\documentclass[11pt,a4paper]{article}
\pdfoutput=1

\usepackage{jheppub}
\usepackage[english]{babel}
\usepackage[utf8]{inputenc}
\usepackage{epsf}
\usepackage{graphicx}
\usepackage{subcaption} 

\usepackage{verbatim} 

\usepackage{amsmath}
\usepackage{amssymb}
\usepackage{mathrsfs}
\usepackage{slashed}

\usepackage{mathbbol}

\usepackage{amsfonts}
\usepackage{bbding}
\usepackage{array}
\usepackage{lineno}
\usepackage{soul}

\newcommand{\y}{y_s^{\alpha}}
\newcommand{\TRH}{T_{\rm rh}}
\newcommand{\TFO}{T_{F}^{\rm fo}}
\newcommand{\AL}[1]{\textcolor{orange}{#1}}

\usepackage{color} 
\usepackage{tikz}
\usetikzlibrary{shapes,arrows,positioning,automata,backgrounds,calc,er,patterns}
\usepackage[compat=1.1.0]{tikz-feynman}
\vspace*{1cm}
\allowdisplaybreaks
\title{Freeze-in at all couplings}

\author[1]{Andreas Goudelis}
\author[2]{\!\!, Andre Lessa}
\author[2]{\!\!, Lucas Magno Dantas Ramos}
\author[1]{\!\!, Thomas Reggio}

\affiliation[1]{Laboratoire de Physique de Clermont Auvergne (UMR 6533), CNRS/IN2P3, Univ.\ Clermont Auvergne, 4~Av.\ Blaise Pascal, F-63178 Aubi\`ere Cedex, France}
\affiliation[2]{
Instituto de F\'isica, Universidade de S\~ao Paulo, 05315-970 S\~ao Paulo, Brazil}

\abstract{
We perform a comprehensive analysis of a charged parent freeze-in dark matter model, focusing on scenarios where the Universe reheats to a temperature comparable to or lower than the mass scales of the theory. In such configurations, dark matter production is Boltzmann-suppressed, allowing for stronger couplings between dark matter and the Standard Model thermal bath while still reproducing the observed relic abundance. 
We emphasize the non-trivial interplay between the reheating temperature, the mediator and dark matter masses and the coupling strength. We show that tracking the number density evolution of both dark matter and the mediator is essential to obtain reliable predictions, including unexpected behaviors such as the mediator non-equilibration due to fast decays. Lastly, we explore the phenomenological implications of this scenario, updating constraints from LHC searches and lepton flavour-violating decays and highlighting the complementarity of these searches in probing the cosmologically viable parameter space.
}
\begin{document}
\maketitle

\section{Introduction}\label{sec:intro}

Since several decades, the question of the microscopic nature of dark matter (DM) has emerged as a major driver for developments in high-energy physics, cosmology and astrophysics: it has lead to the construction of a large number of theories Beyond the Standard Model (BSM) of particle physics, the critical (re-)examination of several assumptions concerning the thermal history of the Universe, the discovery and/or study of a plethora of astrophysical effects and, equally important, the deployment of a massive worldwide experimental campaign in order to address the question ``what is dark matter?'' -- for a recent review \textit{cf e.g.} \cite{Cirelli:2024ssz}.

One of the main questions that any dark matter model aims to answer is why there is as much dark matter in the Universe as is inferred from CMB observations \cite{Planck:2018vyg}. Among the different ideas that have been proposed in order to do so the so-called ``freeze-in'' mechanism \cite{McDonald:2001vt,Hall:2009bx} has, since more than two decades, occupied an increasingly pivotal place. In the freeze-in picture, dark matter was effectively absent in the very early Universe and was produced via the decay or annihilation of bath particles occurring at a sub-Hubble rate. 

In typical freeze-in scenarios in which the radiation domination era began at a high enough temperature\footnote{In this context, ``high enough'' means higher than all mass scales and other relevant temperatures in the theory -- the latter remark will become clearer later on.}, these low reaction rates are attributed to the fact that the interactions between dark matter and the visible sector are extremely weak (``feeble''). Arguably, this was one of the reasons that made freeze-in an attractive possibility in the first place, due to the fact that in many models it becomes possible to evade the stringent constraints stemming from direct, indirect and collider searches for dark matter. The price to pay is that, modulo a few notable exceptions \cite{Hambye:2018dpi, Heikinheimo:2018duk}, it may become \textit{too} difficult to detect dark matter.

However, feeble couplings are not a prerequisite for freeze-in. For instance, if following inflation the Universe reheated to a sufficiently low temperature, dark matter production may have been Boltzmann-suppressed from the start \cite{Belanger:2018sti, Brooijmans:2020yij}. In such scenarios, and depending on the ratio between the available temperatures and the relevant mass scales characterizing the theory, successful freeze-in may be attained for fairly strong interactions between dark matter and the visible sector \cite{Cosme:2023xpa,Brooijmans:2020yij,Arcadi:2024wwg,Belanger:2024yoj,Boddy:2024vgt,Barman:2024nhr,Barman:2024tjt,Borah:2025ema,Arias:2025tvd,Bertou:2026osq,Ghosh:2026mda,Cerdeno:2026lgw}. This opens up an entire realm of possibilities interpolating between traditional freeze-in and thermal freeze-out, with important phenomenological consequences.

In this work we perform a detailed analysis of the cosmology and the phenomenology of such a low reheating temperature ($\TRH$) scenario, focusing on the case of a singlet scalar dark matter candidate ($s$) interacting through a heavy, vector-like lepton mediator ($F$) with the Standard Model leptons.
Some first steps in this direction were already taken in \cite{Belanger:2018sti,Brooijmans:2020yij}, in which most of the emphasis was put on the collider phenomenology of such a model and, consequently, on the relation between $\TRH$ and the mass $m_F$ of the heavy fermions whereas the dark matter mass was taken to be very low (in the keV - MeV range). In this paper we substantially expand upon these original analyses: first, we consider more possibilities for the reheating temperature while allowing the dark matter mass to span a larger range -- in particular, we consider the possibility that $\TRH$ be lower than \textit{both} $m_F$ and $m_s$. As we will see, this can lead to a dramatic increase of the required coupling in order to explain the observed dark matter abundance in the Universe, with important phenomenological consequences. We moreover pay particular attention to the number density evolution of the heavy mediator, which we find to behave non-trivially across the available parameter space, and to its interplay with the predicted dark matter abundance.
Finally, we obtain updated constraints on this scenario from LHC searches for prompt and long-lived signatures as well as constraints from flavour observables.

All in all, our findings illustrate the fact that, much like other cosmic DM production mechanisms, freeze-in does not necessarily point to specific corners of the parameter space of concrete microscopic models: depending on the underlying cosmological assumptions, the observed dark matter abundance in the Universe can be explained through wildly different values of the underlying masses and interactions strengths. This, in turn, implies that dark matter searches probe combinations of microscopic models and cosmological scenarios (and, hence, DM cosmological production mechanisms).

The paper is organised as follows: in Section \ref{sec:themodel} we describe the model and discuss the different contributions to the predicted dark matter abundance. In Section \ref{sec:pheno} we present the phenomenological constraints that the model is subject to, along with some of the most promising channels for the detection of New Physics signals in different experiments. In Section \ref{sec:results} we present our main results. Lastly, in Section \ref{sec:conclusions} we summarize our findings and conclude.

\section{Dark matter in a charged parent freeze-in model}\label{sec:themodel}

\subsection{The model}

The model that we consider is an extension of the Standard Model (SM) by a real, gauge-singlet scalar field $s$ (which will play the role of our dark matter candidate) along with a vector-like fermion $F$ which is a singlet under $SU(2)_L$ and $SU(3)_c$, but charged under $U(1)_Y$. With these charge assignments, $F$ corresponds to a vector-like (charged) lepton. Furthermore, we introduce a discrete $\textbf{Z}_2$ symmetry under which the SM fields are even whereas $s$ and $F$ are odd. Thus, the mass hierarchy $m_s < m_F$ ensures that the DM candidate is stable. 

Under these symmetry assumptions, the most general renormalizable Lagrangian that can be written for the given particle content reads
\begin{align}\label{eq:lag}
    \mathcal{L} & = \mathcal{L}_{SM} + \partial_{\mu}s
\partial^{\mu}s - \frac{\mu_s^2}{2}s^2 +\frac{\lambda_s}{4}s^4 +\lambda_{sh}\,s^2 \left( H^{\dagger}H\right) \\ \nonumber
& + \bar{F}\left(i
\slashed{D}
\right)F - m_F\bar{F}F - \sum_{\alpha=e,\mu,\tau}\y\left(s\bar{F}l_R^\alpha + h.c.\right)
\end{align}
where $\mathcal{L}_{SM}$ is the SM Lagrangian and $l_R^\alpha = \left(e_R,\mu_R, \tau_R \right)$ are the SM right-handed leptons. After electroweak symmetry breaking the dark matter candidate acquires a mass $m_s^2 = \mu_s^2 + \lambda_{sh}v^2$, while the vector-like fermion mass is directly given by $m_F$.

The dark matter candidate can communicate with the visible sector through two distinct ways: the Higgs portal term $\propto \lambda_{sh}$ and the Yukawa-like interactions $\propto \y$. In the limit $\y = 0$ and $\lambda_{sh} \neq 0$, the model behaves effectively much like the singlet scalar dark matter model \cite{Burgess:2000yq,McDonald:2001vt}, which has been extensively studied in the literature -- including the case of a low reheating temperature \cite{Cosme:2023xpa,Arcadi:2024wwg}\footnote{Note, however, that with the given particle content this limit would be problematic since the heavy, charged fermions $F$ would also be cosmologically stable.}. Throughout the subsequent analysis we will, rather, focus on the Yukawa-type interaction induced by the last term of the Lagrangian \eqref{eq:lag} and place ourselves in the opposite limit $\lambda_{sh} = 0$, $\y \neq 0$.

\subsection{The dark matter relic density}\label{sec:relicgeneral}

Under the above assumptions, the dark matter phenomenology of the model depends strongly on the magnitude of the $\y$ couplings. Consider first the conventional case in which, following inflation, the Universe reheated to a very high temperature, much higher than the masses of all particles in the model. We will henceforth refer to this case as the ``effectively infinite reheating temperature'' (IRT) scenario. If at least one of the $\y$ couplings is sufficiently large, then at high cosmic temperatures dark matter equilibrates with the SM thermal bath -- as does the vector-like mediator, given that it is charged under the SM gauge group. In this scenario, the singlet $s$ freezes-out as in usual WIMP dark matter models.

In the opposite regime of very small $\y$, and still sticking for the moment to the IRT scenario, the interactions between $s$ and the SM thermal bath (which, as before, also includes the vector-like leptons $F$) are not strong enough for the two sectors to equilibrate. In this feebly-coupled regime, the final dark matter abundance receives two distinct (in the sense of the relevant temperatures involved) contributions: \textit{i}) a freeze-in contribution which can stem either from $F$ decays while it is still in equilibrium or from annihilations of all bath particles and \textit{ii}) a superWIMP-type \cite{Feng:2003xh} contribution from decays of the heavy fermions $F$ after they have frozen out. To be noted is the fact that in the model that we consider, and for a given value of $m_F$ and $m_s$, the latter contribution is \textit{fixed}, since $F$ only communicates with the SM thermal bath through gauge interactions. 
Ignoring, for the moment, such superWIMP-type contributions as well as dark matter annihilations (``backreactions''), in this regime the dark matter relic abundance is given by
\begin{align}\label{eq:BoltzmannEqs}
Y_s^0 = \int_{T_{\rm CMB}}^{T_{\rm rh}} \frac{dT}{T H(T) {\rm s}(T)} \sum_i \left[ C_i {\cal{N}}_i({\rm bath} \rightarrow sX)_i\right]
\end{align}
where $Y_s^0 = n_s/{\rm s}$ is the dark matter yield, $T$ is the temperature, $H$ is the Hubble expansion rate\footnote{At this stage we have ignored temperature variations of the entropy effective degrees of freedom, \textit{cf} the discussion in \cite{Belanger:2018ccd}.}, ${\rm s}$ is the entropy density (which we assume to be dominated by the SM thermal bath), $C_i \equiv 2$ ($1$) if $X = (\neq) s$, $\mathcal{N}_i$ is the integrated collision term quantifying the number of each $({\rm bath} \rightarrow sX)_i$ reactions taking place per unit space-time volume, and the sum runs over all processes leading to dark matter production.
Typically, in the IRT regime the leading contributions to the dark matter relic density turn out to come from $F$ decays, as long as the process $F \rightarrow s + l^\alpha$ is kinematically allowed. 
\\
\\
A few comments are in order:
\begin{itemize}
    \item Assuming that no DM production took place during reheating and that the latter was instantaneous, equation \eqref{eq:BoltzmannEqs} involves an integration ranging from the temperature of CMB formation $T_{\rm CMB}$ up to the reheating temperature $\TRH$.
    \item In the case of decay processes $F \to s + l^\alpha$, the dark matter production rate scales as $n_F \times \Gamma_F^{s}$, where $n_F$ is the mediator number density and $\Gamma$ its decay width. On the other hand, for annihilation processes of the type $b_1 + b_2 \to s + s$, where $b_i$ represent bath particles, the dark matter production rate scales as $\sqrt{n_{b_{1}} n_{b_{2}}} \times \left\langle \sigma v \right\rangle$, where $\left\langle \sigma v \right\rangle$ is the thermally averaged annihilation cross-section for the reaction in question. 
\end{itemize}
Given these remarks, it becomes clear that dark matter freeze-in does \textit{not} necessarily require feeble couplings. It necessitates sub-Hubble dark matter \textit{production rates}. Suppressed production rates could be due either to the feeble nature of the underlying interactions (\textit{i.e.} small $\Gamma$ and/or $\left\langle \sigma v \right\rangle$) or, which is the case of interest in this paper, to the fact that the relevant number densities impacting dark matter production are suppressed (\textit{i.e.} small $n_F$ and/or $n_{b_{i}}$), \textit{cf e.g.} \cite{Belanger:2018sti, Cosme:2023xpa, Arcadi:2024wwg}.

From the previous discussion, we can understand the qualitative features of what is expected to happen if, following inflation, the Universe only reheated to a temperature which is comparable with or lower than (some of) the mass scales relevant for dark matter production. Concretely, in the model under consideration:
\begin{itemize}
    \item once $\TRH < m_F$, and as long as the mediator is not decoupled, its number density becomes Boltzmann-suppressed. This, in turn, implies that once $n_F(\TRH)$ becomes sufficiently low, the contribution from $F$ decays may become subleading and SM particle annihilation processes may become dominant. \footnote{With a bit of hindsight, let us point out that in practice we find that only annihilations involving SM leptons in the initial state are relevant in cosmologically viable regions of the parameter space.}
    \item Moreover, once $\TRH < m_s$, dark matter production from the thermal bath (be it from $F$ decays or from bath particle annihilations) becomes Boltzmann-suppressed \textit{altogether}. 
    \item As an additional point, once $\TRH$ becomes smaller than the approximate $F$ freeze-out temperature $T_F^{\rm fo}$, it is no longer reasonable to consider that the vector-like leptons freeze-out, since they never equilibrate with the SM thermal bath. Instead, a more reasonable assumption is to rather consider that they freeze-in, with the resulting $F$  population eventually decaying into $s$ states.
\end{itemize}
In summary, as long as equilibrium is never established between $s$ and the SM, and depending on the hierarchy between $m_s$, $m_F$, $T_F^{\rm fo}$ and $\TRH$, dark matter can be produced through four different mechanisms: \textit{i}) decays of in-equilibrium $F$'s, \textit{ii}) annihilations of SM particles, \textit{iii}) decays of $F$'s after they freeze-out and \textit{iv}) decays of frozen-in $F$'s. In order to take all these processes into account, it is necessary to jointly track the number density evolution of both $s$ and $F$, \textit{i.e.} solve a system of coupled Boltzmann equations. Focusing on the leading terms, these read
\begin{align}\label{eq:boltzmannspecific}
    \frac{dY_s}{dx} & = \frac{1}{3H} \left|\frac{d{\rm s}}{dx} \right| \left[ \langle\sigma v\rangle_{s s} \left( {\bar{Y}_s}^2 - Y_s^2\right) + \langle\sigma v\rangle_{sF} \left( \bar{Y}_s \bar{Y}_F - Y_s Y_F\right) + \frac{\Gamma}{\gamma {\rm s}} Y_F \right] \\ \nonumber
    \frac{dY_F}{dx} & = \frac{1}{3H} \left|\frac{d{\rm s}}{dx} \right| \left[ \langle\sigma v\rangle_{F F} \left( {\bar{Y}_F}^2 - Y_F^2\right) + \langle\sigma v\rangle_{sF} \left( \bar{Y}_s \bar{Y}_F - Y_s Y_F\right) - \frac{\Gamma}{\gamma {\rm s}} Y_F \right]
\end{align}
where $x \equiv m_s/T$, $\bar{Y}_i$ represents the equilibrium yield for the species $i$, $\langle \sigma v \rangle_{ij}$ is the thermally averaged annihilation cross-section for thermal bath particles into $i+j$,  $\gamma = K_2 (m_F/T)/K_1(m_F/T)$ is the relativistic factor and ${\rm s}$ the entropy density.

In what follows we will make use of the {\tt FeynRules} \cite{Alloul:2013bka} implementation of the model which was presented in \cite{Belanger:2018sti}. The computation of the different yield evolutions, along with the DM relic density, is performed with {\tt CalcHEP/micrOMEGAs 6} \cite{Belyaev:2012qa,Alguero:2023zol}, which allows the simultaneous resolution of the coupled system of Boltzmann equations\footnote{Note that the full-blown Boltzmann equations contain additional terms beyond the ones presented in Eq.\eqref{eq:boltzmannspecific}, most notably terms corresponding to conversion processes \cite{Garny:2017rxs,DAgnolo:2017dbv}. Although these \textit{are} taken into account in our numerical analysis, they are largely subleading and we have omitted them in Eq.\eqref{eq:boltzmannspecific} for the sake of brevity.}. Throughout our analysis we will assume that reheating was instantaneous, that only the SM thermal bath was produced upon inflaton decays and that kinetic equilibrium holds at all times between all sectors. However, we go beyond the conventional freeze-in approximation, in that we also keep track of both $F$ and $s$ backreactions. The latter allows us, in particular, to correctly capture the reheating temperature value for which $F$ equilibrates with the plasma. Lastly, in everything that follows we will choose for simplicity a universal value for the $y_s^\alpha$ couplings across lepton families, $y_s^\alpha \equiv y_s \ \forall \ \alpha = e, \mu, \tau$.

\subsubsection{The mediator abundance}\label{sec:themediator}

Given the importance of the mediator $F$ for dark matter production, it is instructive to first examine the general behaviour of their number density evolution.

\begin{figure}
    \centering
    \includegraphics[width=0.8\linewidth]{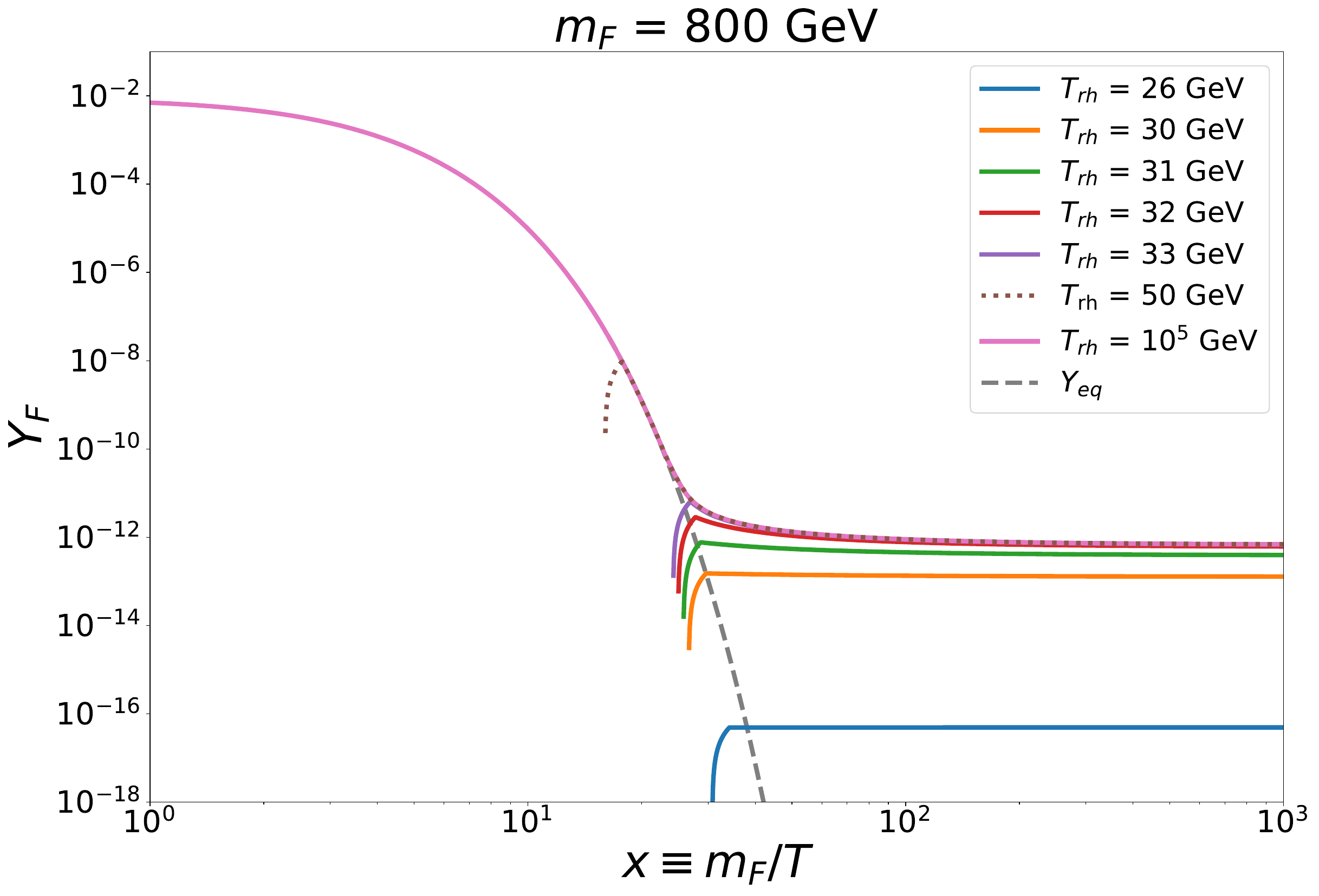}
    \caption{Temperature evolution of the vector-like lepton yield ($Y_F$) ignoring decays into $s$, for $m_F = 800$ GeV and different values of the reheating temperature.}
    \label{fig:fevolution}
\end{figure}

Let us first ignore $F$ decays into dark matter. For a benchmark mass $m_F = 800$ GeV, we fix the initial mediator yield $Y_F(\TRH)$ to zero and compute its subsequent evolution for different values of the reheating temperature ranging from the IRT regime down to values well below the $F$ freeze-out temperature $T_F^{\rm fo} \simeq 30$ GeV
\footnote{We should note at this point that the mediator freeze-out is a smooth process, hence $T_F^{\rm fo}$ only indicates the approximate temperature where the freeze-out process starts.
In this sense, and as we will see, all transitions between different regimes are gradual. Therefore, in the following we will refer to the (hypothetical) $F$ freeze-out temperature in the absence of $F$ decays as the ``na\"ive'' freeze-out temperature.}.
Our numerical results are shown in Figure~\ref{fig:fevolution}.

As we see, for reheating temperatures well above $T_F^{\rm fo}$, the mediator abundance rapidly reaches its equilibrium value and subsequently freezes-out.
For slightly smaller reheating temperatures, 30 GeV $\lesssim \TRH \lesssim 32$ GeV, 
we observe an initial increase and a subsequent reduction of the abundance of $F$ which follows a freeze-out-like evolution but eventually settles to a smaller abundance compared with the one that would be achieved through conventional freeze-out.
This behaviour, which is due to the fact that $F$ production and annihilation take place at slightly different temperatures, has already been noted in \cite{Cosme:2023xpa}.
Finally, if the reheating temperature is further reduced, $\TRH < 30$ GeV, $Y_F(T)$ follows a fairly standard freeze-in evolution due to annihilations of thermal bath (SM) particles, as shown by the blue curve in Figure~\ref{fig:fevolution}.

\begin{figure}
    \centering
    \includegraphics[width=0.8\linewidth]{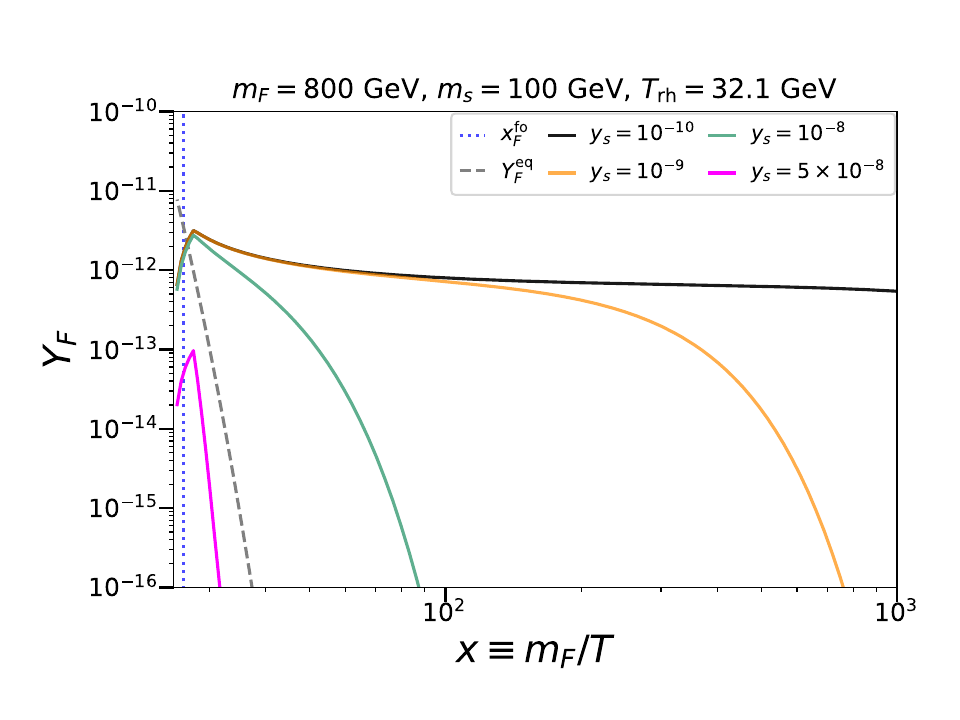}
    \caption{Temperature evolution of the vector-like lepton yield ($Y_F$) taking into account decays into $s$, for a fixed value of the reheating temperature $\TRH = 32.1$ GeV and for different values of the $y_s$ coupling, assuming $m_F = 800$ GeV and $m_s = 100$ GeV.}
    \label{fig:fequilibration}
\end{figure}

Let us now turn on $F$ decays into dark matter. In Figure \ref{fig:fequilibration} we show the temperature evolution of $Y_F$ assuming $m_F = 800$ GeV, $m_s = 100$ GeV\footnote{This choice is less relevant as long as the decay $F \rightarrow s + l^\alpha$ is not phase space-suppressed.}, $\TRH = 32.1$ GeV (\textit{i.e.} a regime which lies at the verge of na\"ive $F$ equilibration, according to the findings of Figure \ref{fig:fevolution}) and for different values of the coupling $y_s$ (solid lines). The gray-dashed line depicts $Y_F^{\rm eq}$ whereas the blue-dotted vertical line displays the would be freeze-out temperature of $F$ if its decays into DM are ignored.

As we observe, an interesting feature appears: even though both the $F$-SM interaction strength (gauge interactions) and the fraction of the plasma which can lead to mediator production (which depends on $m_F/\TRH$) are fixed, the equilibration between the two sectors also depends on the interaction strength between $F$ and $s$. Concretely, once $y_s$ becomes sufficiently large, $F$ never attains its equilibrium abundance. This is due to the fact that $F$ decays become sufficiently fast such that the reaction $FF \rightarrow {\rm SM \ SM}$ becomes suppressed, due to the continuous (fast) reduction of $n_F$. Put simply, the mediator decays faster into DM than it can be produced by annihilations of the thermal bath particles and $F$ never reaches thermal equilibrium.

From this discussion it becomes clear that the evolution of the $F$ abundance is highly non-trivial and needs to be kept track of. As we will see in the following Sections, whether or not it impacts the predicted dark matter abundance depends on the region of parameter space under consideration and more precisely \textit{i}) on the relation between $\TRH$ and the two mass scales of the theory and \textit{ii}) when $\TRH \sim T_F^{\rm fo}$, on the magnitude of $y_s$. In practice, in everything that follows when computing the dark matter abundance the initial abundances of $F$ and $s$ will be set to zero, unless $\TRH \gg T_F^{\rm fo}$. In the latter case it will be set to $\bar{Y}_F$, a choice which is dictacted by numerical stability reasons.
\subsubsection{Putting the pieces together}\label{sec:alltogether}

\begin{figure}
    \centering
    \includegraphics[width=0.8\linewidth]{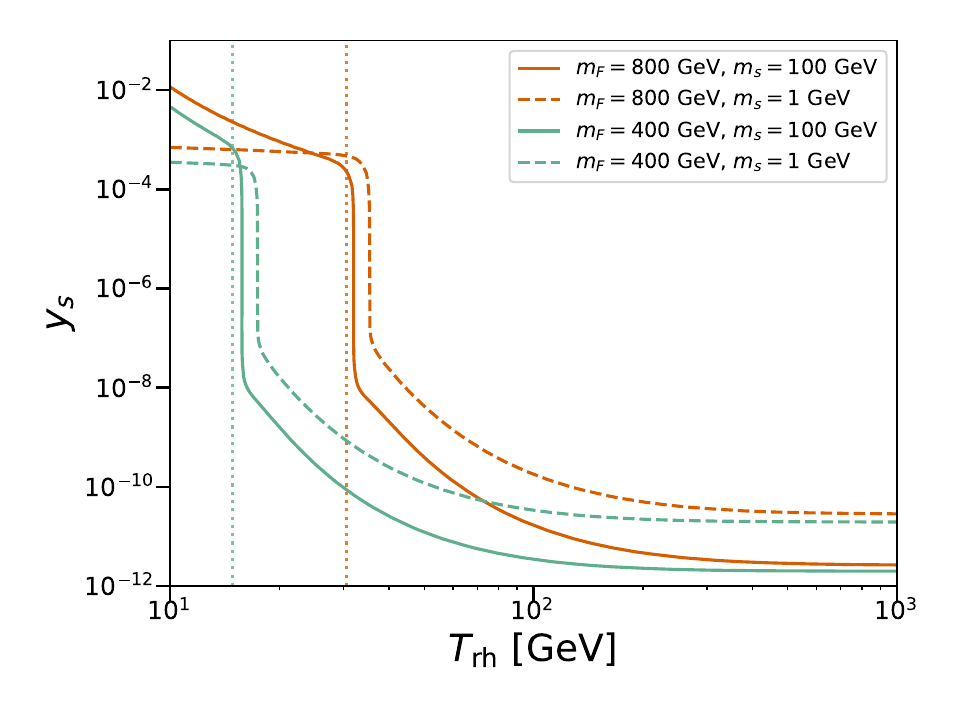}
    \caption{Contours of $\Omega h^2 = 0.12$ in the $(\TRH, y_s)$ plane for four representative benchmarks $m_F = 800/400$ GeV (red/green lines) and $m_s = 100/1$ GeV (solid/dashed lines). The left/right-most dotted lines indicate the (approximate) $F$ freeze-out temperature for $m_F = 400/800$ GeV respectively.}
    \label{fig:TRyContours}
\end{figure}

A detailed numerical analysis of the cosmologically viable parameter space will be presented in Section \ref{sec:results}. However, we can already illustrate some general aspects of its behaviour as a function of the reheating temperature. To this goal, in Figure \ref{fig:TRyContours} we show contours of $\Omega h^2 = 0.12$ in the $(\TRH, y_s)$ plane for four representative benchmarks $(m_F, m_s) = (800, 100)$, $(800,1)$, $(400,100)$ and $(400,1)$ GeV (red solid, red dashed, green solid and green dashed lines, respectively) resulting from the solution of the coupled set of Boltzmann equations as described previously.

The general behaviour of these results can be understood through the previous discussion. For large enough values of $\TRH$, and for a given mass combination, a more or less constant (feeble) value of the Yukawa-like coupling is required in order to satisfy the relic density constraint. This is the IRT regime, in which the predicted dark matter abundance becomes insensitive to the precise value of $\TRH$. In this limit, the leading contribution to $Y_s$ comes from $F$ decays while they are still in equilibrium, along the lines of conventional freeze-in. Note that for the given mass combinations, the superWIMP contribution turns out to be subleading. In Section \ref{sec:results} we will see examples in which this contribution becomes dominant, in particular for larger values of $m_s$.

When $\TFO \lesssim  \TRH \lesssim m_F$ the abundance of $F$ becomes Boltzmann-suppressed, despite still reaching thermal equilibrium; a corresponding increase in $y_s$ is required to compensate this effect. However, as the reheating temperature starts approaching the (na\"ive) $F$ freeze-out temperature, two effects come into play: first, the capacity to produce \AL{$F$} from plasma scattering processes becomes even further suppressed due to the decrease of $\TRH/m_F$. 
In principle, this effect could be compensated by increasing $y_s$. 
However, such an increase in $y_s$ leads to a further suppression of $F$, which can no longer reach thermal equilibrium due to its fast decay rate, as discussed in Section \ref{sec:themediator}.
As a result, the dark matter production through decays becomes suppressed and the production through scatterings of thermal plasma leptons starts becoming dominant.

The sharp increase of $y_s$ seen  around $\TRH \sim \TFO$ can be understood as the transition from the decay-dominated to the scattering-dominated regime.
While in the (decay-dominated) IRT regime, the dark matter yield scales as $Y_{\rm dec} \propto y_s^2/m_F$, in the scattering-dominated regime the DM production cross-section scales roughly as $\langle \sigma v \rangle \sim y_s^4 T^2/m_F^4$.
Given that DM production peaks around the reheating temperature (since $T_{\rm rh} \ll m_F$), the scattering contribution ends up scaling as $Y_{\rm scat} \propto y_s^4 T_{\rm rh}^3/m_F^4$. Therefore, in order to reproduce the observed relic abundance within the scattering-dominated region ($\TRH < \TFO$) one obtains $y_{s}^{\rm scat} \sim {\cal{O}}(10) \times \sqrt{y_{s}^{\rm dec}}$. This corresponds fairly well to the behaviour of the dashed lines observed in Figure \ref{fig:TRyContours}, in which the required coupling in the scattering-dominated regime (left-most part of the curves) increases by roughly 7 orders of magnitude with respect to the IRT one (right-most part of the curves). Besides, in the transition regime (quasi-vertical part of the curves), the DM relic density is driven by a combination of decay and scattering processes, with the latter rapidly taking over as $T_{\rm rh}$ decreases.

Lastly, for even smaller values of $\TRH$ ($\TRH < \TFO$), (freeze-in) dark matter production is purely driven by annihilation processes. In this regime, if $m_s \ll \TRH$, the dark matter production is independent of the reheating temperature (and, of course, the $F$ abundance), depending only on $y_s$. This explains the behaviour of the dashed lines seen in Figure \ref{fig:TRyContours}
for $m_s = 1$~GeV and $\TRH < \TFO$. However, if $m_s \gtrsim \TRH$, the dark matter production suffers an additional Boltzmann suppression, requiring even larger couplings when compared to the light $m_s$ scenario. This kinematical suppression is enhanced as $\TRH$ decreases, thus requiring an increase in $y_s$ to reproduce the observed dark matter relic abundance.

Let us stress that the discussion above is based on two assumptions for the reheating mechanism: {\it i)} it is instantaneous and {\it ii)} it is consistent with the initial conditions of zero abundances for the mediator and dark matter.  As long as the first assumption is satisfied, relaxing {\it ii)} would typically introduce an additive contribution to the RHS of Eq.\eqref{eq:BoltzmannEqs} which would depend on the inflaton ``branching fractions''. The impact of a finite-duration reheating period can be more delicate: quantifying its effect requires specifying the inflationary model, its duration and the maximum temperature of the Universe ($T_{\rm max}$) achieved. Dark matter production during reheating has, \textit{e.g.}, been recently discussed in \cite{Belanger:2024yoj,Bertou:2026osq}.
Nonetheless these effects can be approximately parametrized by using an ``effective reheating temperature'' with a value between $T_{\rm max}$ and the thermal bath temperature at the onset of radiation domination, and which in general can depend non-trivially on all parameters of the underlying reheating scenario. For simplicity in this work we refer to this effective reheating temperature as $T_{\rm rh}$.

In the following Sections we will discuss in much more detail how distinct choices of $m_F$, $m_s$, $\TRH$ and $y_s$ can impact the interplay between different contributions to the dark matter relic density and existing experimental constraints -- which, as we will see, can become rather non-trivial as one moves across the parameter space.
\section{Phenomenological constraints}\label{sec:pheno}

Let us now move to the phenomenological constraints that the model is subject to. These stem from three distinct sources: {\it i)} direct searches, {\it ii)} searches for lepton flavour-violating decays and {\it iii)} searches for production and decays of the mediator at colliders.

\subsection{Direct detection}\label{sec:DD}

Since the model we consider does not involve tree-level interactions between dark matter and quarks/gluons, we expect the constraints stemming from searches for dark matter - nucleus scattering to be negligible.\footnote{Since the vector-like lepton ($F$) only couples to right-handed leptons, the one-loop coupling to quarks vanishes. As a result, the coupling can only arise at the two-loop level, thus being suppressed by $y_s^2 g'^2/\left(4 \pi\right)^4$.} However, as long as the coupling to first-generation leptons is non-zero, it does lead to dark matter - electron scattering at tree-level. The latter has been targeted by different direct detection experiments including Super-CDMS~\cite{SuperCDMS:2020ymb}, DAMIC~\cite{DAMIC:2019dcn}, SENSEI~\cite{SENSEI:2020dpa}, PandaX-II~\cite{PandaX-II:2021nsg}, DarkSide-50 (DS-50)~\cite{DarkSide:2022knj} and XENON1T-S2 (XE1T-S2)~\cite{XENON:2019gfn}. The corresponding constraints are typically presented in terms of a reference cross-section $\bar{\sigma}_e$ which is defined as \cite{Essig:2011nj}
\begin{equation}
    \bar{\sigma}_e \equiv \frac{\mu_{se}^2}{16^\pi m_s^2 m_e^2} \left. \overline{\left| {\cal M}(q) \right|^2} \right|_{q^2 = (\alpha m_e)^2}
\end{equation}
where $\mu_{se}$ is the dark matter-electron reduced mass, $q^2$ is the momentum transfer and the matrix element ${\cal M}$ is defined through
\begin{equation}
    \overline{\left| {\cal M}(q) \right|^2} = \left. \overline{\left| {\cal M}_{\rm free}(q) \right|^2} \right|_{q^2 = (\alpha m_e)^2} \times \left|F_{DM}(q)\right|^2
\end{equation}
where ${\cal M}_{\rm free}(q)$ is the usual matrix element for dark matter - free electron scattering and $F$ is an appropriate form factor. In our case, this form factor is effectively equal to $1$, \textit{cf} the discussion in \cite{Essig:2011nj}. Our numerical analysis showed that the limits set by searches for dark matter - electron scattering are not yet strong enough to place significant bounds on our model.
For example, assuming $m_s=1$ GeV and $m_F=400$ GeV, only couplings of ${\cal{O}}(10^3)$ or larger are excluded.

\subsection{Flavour constraints}\label{sec:flavour}

An additional constraint on the model arises from searches for radiative flavour-violating charged lepton decays. In our case, the most stringent limit stems from the limit on the $\mu \rightarrow e \gamma$ branching ratio\footnote{Since in this paper we consider universal couplings to all lepton flavours, the constraint from $\mu \rightarrow e \gamma$ is by far the strongest. However, for other choices of $\y$, additional constraints can also arise from searches for $\tau \rightarrow e \gamma$ and $\tau \rightarrow \mu \gamma$.}  
\begin{equation}\label{eq:MEGlim}
    {\rm BR}(\mu \rightarrow e \gamma) < 1.5 \times 10^{-13} \mbox{ (90\% C.L.)}
\end{equation}
from MEG II~\cite{MEGII:2025gzr}. In our model, the partial width for this process is generated by the diagram in Figure~\ref{fig:diagram_mu2egamma} and given by
\begin{equation}\label{eq:muegamma}
    \Gamma(\mu \rightarrow e \gamma) \simeq \frac{16 e^2 (y_s)^4 (m_\mu^2 + m_e^2)(m_\mu^2 -m_e^2)^3 
    \left[ t\left( \left(t-6\right) t + 3\right) + 6t\log t +2\right]^2}{9 (16\pi)^5 m_\mu^3 m_s^4 (t-1)^8}
\end{equation}
where $e \simeq 0.31$, $t\equiv m_F^2/m_s^2$ and we have assumed the mass of the electron and muon to be much smaller than the mass of $s$ and $F$ in the calculation of the loop integrals. This expression has been calculated using {\tt FeynArts} \cite{Hahn:2000kx} to generate the relevant Feynman diagrams,  {\tt FeynCalc} \cite{Mertig:1990an,Shtabovenko:2023idz} to compute the amplitude and {\tt PackageX/FeynHelpers} \cite{Patel:2016fam,Shtabovenko:2016whf} to obtain analytical expressions for the loop integrals. It has been cross-checked against the general formulae presented in \cite{Lavoura:2003xp}.

\begin{figure}
\centering
\begin{tikzpicture}
\begin{feynman}[medium]
            \vertex (a) {$\mu$};
            \vertex[right=45pt of a] (b);
            \vertex[right=25pt of b] (c1);
            \vertex[above=25pt of c1] (cloop);
            \vertex[right=25pt of c1] (c2);
            \vertex[right=45pt of c2] (d) {e};
            \vertex[above=50pt of d] (e) {$\gamma$};
            \diagram* {
                (a) -- [fermion,thick] (b),
                (b) -- [scalar,thick, edge label=s] (c2),
                (c2) -- [fermion,thick] (d),
                (b) -- [quarter left, fermion,thick, edge label=F] (cloop),
                (cloop) -- [quarter left, fermion, thick] (c2),
                (cloop) -- [photon] (e),
            };          
\end{feynman} 
\end{tikzpicture}
    \caption{Feynman diagram for the LFV decay process $\mu\to e \gamma$, generated at one-loop in our model.}
    \label{fig:diagram_mu2egamma}
\end{figure}

The muon decay width is, in turn, given by $\Gamma_\mu = 1/\tau_\mu$ with $\tau_\mu = (2.1969811 \pm 0.0000022)\times 10^{-6}$ sec \cite{ParticleDataGroup:2024cfk}. By combining Eqs.\eqref{eq:MEGlim}, \eqref{eq:muegamma} and the muon lifetime we can compute the corresponding constraints on the parameters of our model.

\subsection{Collider searches}\label{sec:colliders}

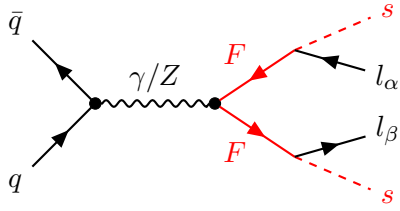
\begin{figure}
\centering
\begin{tikzpicture}[baseline=-0.09cm]
\begin{feynman}[medium]
            \vertex[dot] (a) {};
            \vertex[dot,right=45pt of a] (b) {};
            \vertex[above left= of a] (c) {$\bar{q}$};
            \vertex[below left= of a] (d) {$q$};
            \vertex at ($(b) + (30pt, 20pt)$) (e);
            \vertex at ($(b) + (30pt, -20pt)$) (f);
            
            \vertex at ($(e) + (35pt, 15pt)$) (e1) {\textcolor{red}{$s$}};
            \vertex at ($(e) + (35pt, -10pt)$) (e2)  {$l_{\alpha}$};
            \vertex at ($(f) + (35pt, 10pt)$) (f1)  {$l_\beta$};
            \vertex at ($(f) + (35pt, -15pt)$) (f2) {\textcolor{red}{$s$}};

            \diagram* {
                (a) -- [fermion,thick] (c),
                (d) -- [fermion,thick] (a),
                (a) -- [photon,thick, edge label=$\gamma/Z$] (b),
                (e) -- [red,fermion,thick, edge label'= $F$] (b),
                (b) -- [fermion,thick,red, edge label'= $F$] (f),
                (e) -- [scalar,thick,red] (e1),
                (e2) -- [fermion,thick] (e),
                (f) -- [scalar,thick,red] (f2),
                (f) -- [fermion,thick] (f1),
                
            };          
\end{feynman} 
\end{tikzpicture}
    \caption{Feynman diagram for the production and decay of the mediator at the LHC. Since we assume a universal coupling to all lepton families, $\alpha,\beta= e,\mu,\tau$}.
    \label{fig:diagram_lhc}
\end{figure}

Given the assumption $\lambda_{sh} = 0$ in Eq.\eqref{eq:lag}, the main collider signatures predicted by the model we consider are related with the (Drell-Yan) pair-production of the vector-like leptons and their subsequent decay into SM leptons along with dark matter, as shown in Figure~\ref{fig:diagram_lhc}.
Furthermore, since we assume universal couplings to leptons, the $F$ decays will lead to final states with all possible combinations of opposite sign leptons: $p p \to F^+ + F^- \to l^{\alpha} + s +  l^{\beta} + s$ with $\alpha,\beta = e,\mu,\tau$.

\begin{figure}
    \centering
    \includegraphics[width=0.48\linewidth]{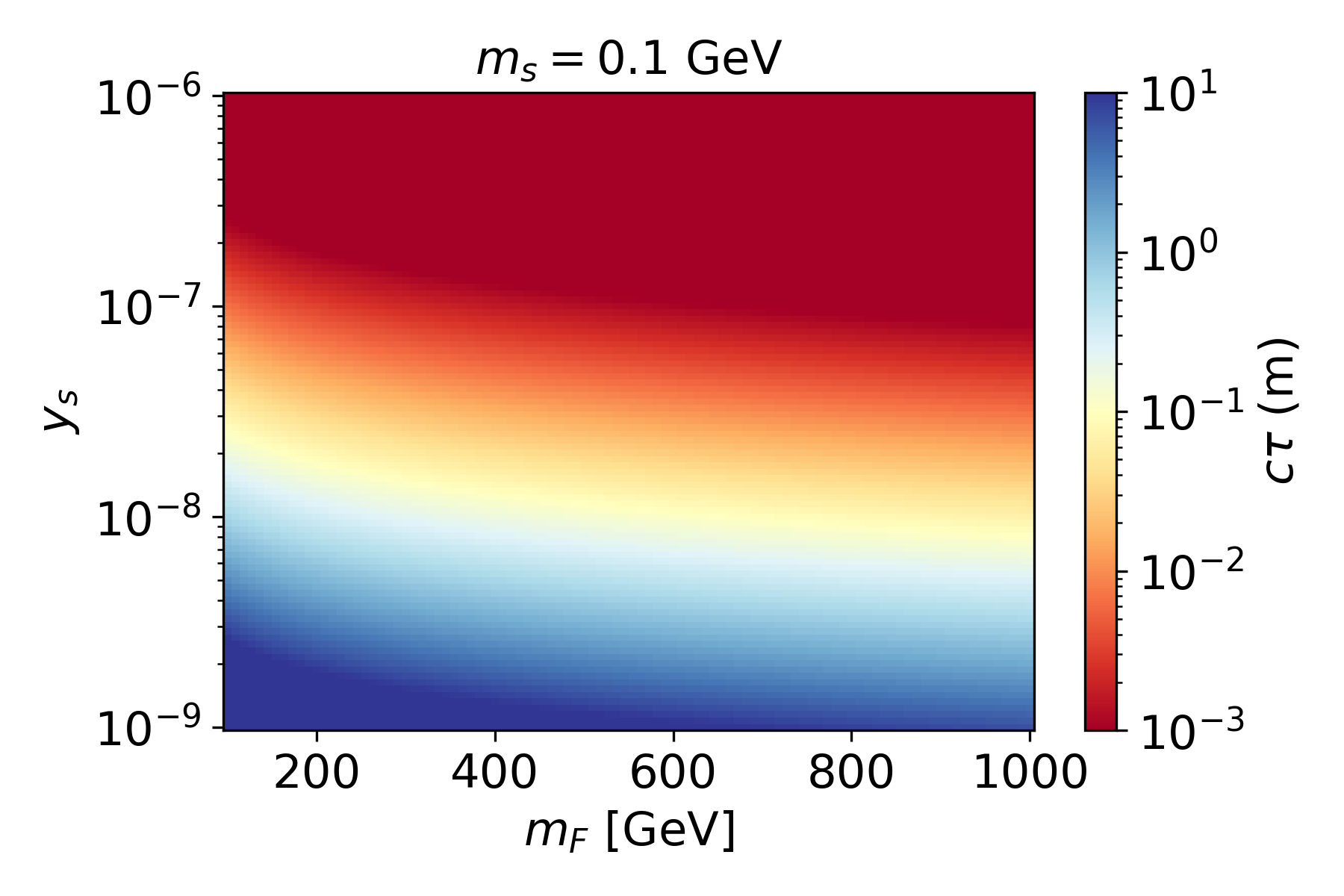}
    \includegraphics[width=0.48\linewidth]{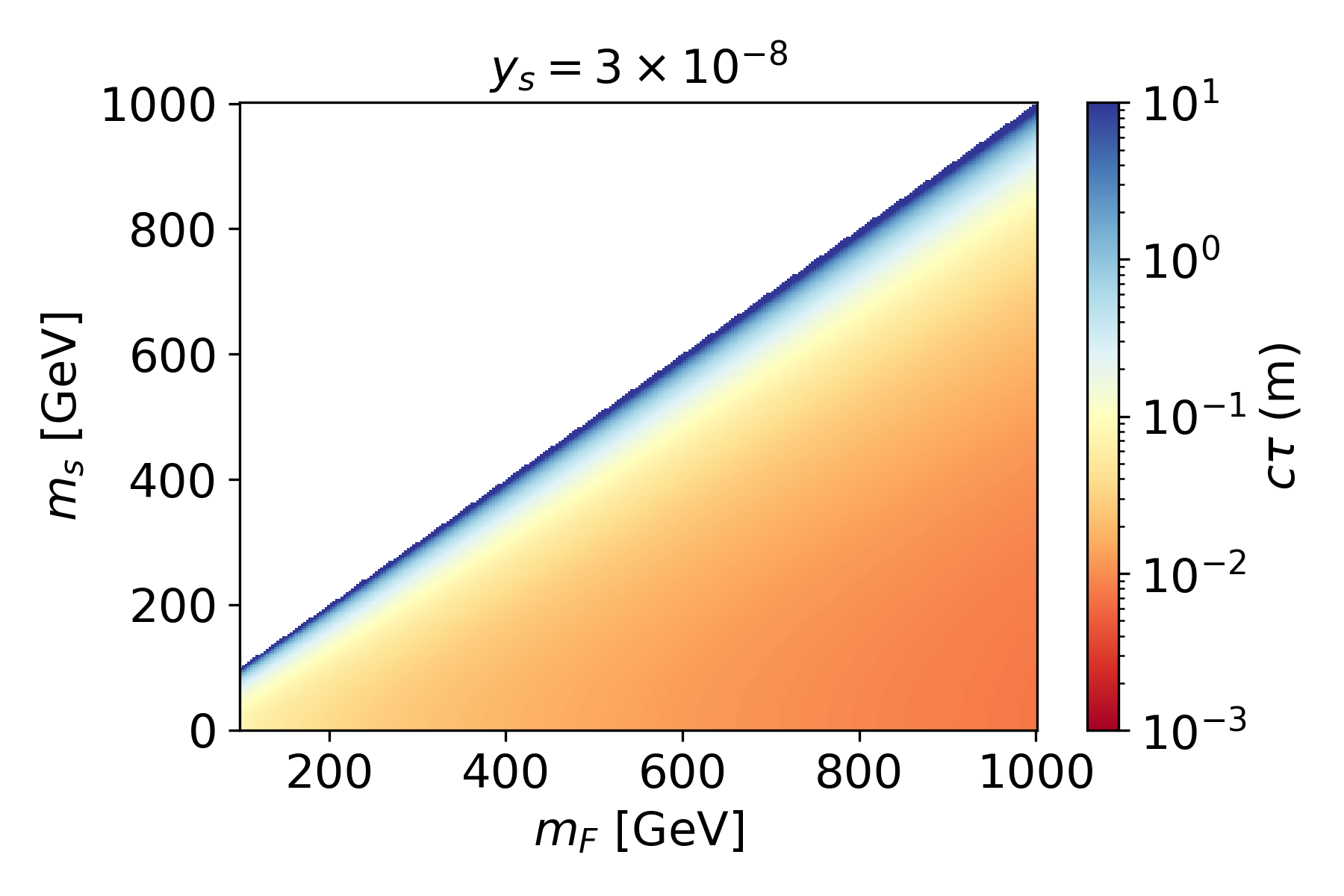}
    \caption{Mediator ($F$) lifetime as a function of the model parameters. {\it Left:} lifetimes for a light $s$ as a function of the coupling $y_s$ and the mediator mass. {\it Right}: lifetimes for a fixed coupling as a function of $m_s$ and $m_F$.}
    \label{fig:lifetimes}
\end{figure}

While the mediator production is fixed by gauge couplings, 
its lifetime is proportional to $1/(y_s)^2$ which, as shown in Figure~\ref{fig:TRyContours}, can span many orders of magnitude depending on the value of $\TRH$. 
In Figure~\ref{fig:lifetimes}a we display the mediator decay length as a function of its mass and $y_s$ for $m_s \ll m_F$. As we can see, the (proper) decay length can be smaller than a millimeter (for $y_s \gtrsim 10^{-7}$) or larger than a meter (for $y_s \lesssim 2\times 10^{-9}$).
Large decay lengths can also be achieved for intermediate coupling values if the decay is phase-space suppressed ($m_s \lesssim m_F$), as shown in Figure~\ref{fig:lifetimes}b.
Therefore, depending on $y_s$, the mediator can decay promptly, within, or even outside the ATLAS and CMS detectors.

Depending on the lifetime of $F$, its decay can lead to distinct long-lived particle (LLP) and/or prompt signatures.
We can schematically distinguish three main regimes~\cite{Belanger:2018sti,Brooijmans:2020yij}:
\begin{itemize}
    \item For $y_s \lesssim \mathcal{O}(10^{-9})$, a sizeable fraction of the produced $F$'s will decay outside the tracker, leading to a track with large $dE/dx$. This signature is the target of searches for Heavy Stable Charged Particles (HSCP).
    \item For intermediate ($10^{-9}$ - $10^{-7}$) coupling values, the mediator will predominantly decay within the tracker, resulting in leptons being produced far from the interaction point. This regime is constrained by searches for displaced leptons (DL). However, if $\Delta m \equiv m_F - m_s \ll m_F$, the leptons become too soft to be reconstructed or triggered on. In this case, disappearing track (DT) searches can become sensitive to the signal. Since the DT searches typically require highly compressed scenarios ($\Delta m \lesssim 1$ GeV) and veto hard leptons, we will not consider them in this work.
    \item Lastly, for $y_s \gtrsim \mathcal{O}(10^{-7})$, most of the produced $F$'s will decay promptly, rendering the model susceptible to constraints from searches for prompt leptons accompanied by large missing transverse energy.
\end{itemize}

In Table~\ref{tab:usedsearches} we list all the LHC searches considered in this work. The last column displays the reinterpretation tool used to compute the constraints.
For all searches the production cross-section for $p p \to F^+ F^-$ was computed at leading order and a universal coupling to all lepton flavors was assumed, \textit{i.e.} $BR(F \to s + l_\alpha) = 1/3$ for $\alpha= e,\mu,\tau$. The only exception is the compressed region with $\Delta m < m_\tau$, where decays into $\tau$'s are kinematically forbidden, enhancing the decays to electrons and muons.

\begin{table}[h!]
  \centering
  \begin{tabular}{c|c|c|c|c}
Signature  & Analysis, Reference & $\sqrt{s}$ [TeV] & $\int\!\!{\cal L}$  [fb$^{-1}$] & Implementation \\
\hline
HSCP & \href{https://cms-results.web.cern.ch/cms-results/public-results/publications/EXO-13-006/index.html}{CMS-EXO-13-006}~\cite{CMS:2015lsu} & 8 & 18.8  & {\tt SModelS}
\\
\hline
HSCP & \href{https://atlas.web.cern.ch/Atlas/GROUPS/PHYSICS/PAPERS/SUSY-2018-42/}{ATLAS-SUSY-2018-42}~\cite{ATLAS:2022pib} & 13 & 139  & {\tt SModelS}
\\
\hline
DL & \href{https://atlas.web.cern.ch/Atlas/GROUPS/PHYSICS/PAPERS/SUSY-2018-14}{ATLAS-SUSY-2018-14}~\cite{ATLAS:2020wjh} & 13 &139 & {\tt LLP recasting}
\\
\hline 
 $2l + E_T^{\rm miss}$ & \href{https://atlas.web.cern.ch/Atlas/GROUPS/PHYSICS/PAPERS/SUSY-2018-32/}{ATLAS-SUSY-2018-32}~\cite{ATLAS:2019lff} & 13 &139  & {\tt MadAnalysis5}
\\
\hline 
\end{tabular}
\caption{Summary of the LHC searches used to constrain the model considered in this work. HSCP stands for Heavy Stable Charged Particle and DL for displaced leptons. The last column displays the reinterpretation tools used to compute the constraints.
}
\label{tab:usedsearches}
\end{table}

The HSCP limits were obtained using {\tt SModelS} v3~\cite{Altakach:2024jwk} and its interface to {\tt micrOMEGAs}.
The database includes efficiency maps for
the 8~TeV CMS search~\cite{CMS:2015lsu} and the Run 2 ATLAS search~\cite{ATLAS:2022pib} for HSCPs, which were used to constrain the parameter space with large lifetimes.
Since the efficiencies are provided as a function of the  LLP mass and width, they allow us to determine the loss in sensitivity for mediator decay lengths under a few meters.

For intermediate lifetimes, $c\tau \sim \mathcal{O}({\rm mm-cm})$, the most relevant search becomes the displaced lepton search by ATLAS~\cite{ATLAS:2020wjh}. The search looks for two displaced leptons (electrons or muons) and considers three signal regions (SRs) targeting all possible flavor combinations: $ee$, $\mu\mu$ and $e\mu$. Therefore, it is well-suited for testing the scenario considered here, where $F$ has universal lepton couplings and can decay to all possible flavor combinations. Note that the $F \to s + \tau$ decays can also contribute to the signal, if the $\tau$'s decay leptonically. The search results have been interpreted in terms of simplified models with long-lived sleptons ($\tilde{l}$) decaying into Gravitinos ($\tilde{G}$) with $m_{\tilde{G}} \simeq 0$. Therefore, the search limits can only be directly applied to the $m_s \to 0$ case. However, a recasting of this search is available in the {\tt LLP Recasting Repository}~\cite{llprecastingRepo} and it was used to extrapolate the limits to higher $m_s$ values.
In order to obtain the constraints for the model considered here we have generated Monte Carlo (MC) events using  
{\tt MadGraph5\_aMC@NLO}~3.6~\cite{Alwall:2014hca} with the NNPDF set~\cite{Ball:2012cx}, interfaced to {\tt Pythia} 8.3~\cite{Bierlich:2022pfr} 
for showering and hadronization.
To properly describe the $p_T$ distribution of the LLPs, events with one additional jet were generated at the matrix element level and the MLM prescription was used for jet-parton matching with a matching scale $q = m_F/4$. 
For each set of parameters, between 30k-50k MC events have been generated.
The reinterpretation of the displaced lepton search available at the  {\tt LLP Recasting Repository} makes use of the parametrized reconstruction efficiencies provided as auxiliary material by ATLAS~\cite{ATLAS:2020wjh}, which can be applied to the truth-level leptons in the generated MC events
\footnote{We have verified that, following the same procedure for the slepton simplified models considered by the analysis, we were able to reproduce the ATLAS efficiencies and exclusion curves within $\sim 20-50$\% for most of the relevant parameter space. More details concerning the validation of the implementation can be found in the LLP repository~\cite{llprecastingRepo}.}.
Using the predicted number of signal events for each signal region, limits were computed using all 3 signal regions and the {\tt Pyhf}~\cite{Heinrich:2021gyp} statistical model provided by ATLAS along with the {\tt Spey}~\cite{Araz:2023bwx} software package and its {\tt Pyhf} plugin.

Once the mediator decay length becomes smaller than a few millimeters, the prompt limits are applicable. Note that the di-lepton search considered here~\cite{ATLAS:2019lff} 
requires the reconstructed leptons to originate from the primary vertex through   cuts on the transverse ($d_0$) and longitudinal ($z_0$) impact parameters: $|d_0|/\sigma(d_0) < 3(5)$ for electron (muon) tracks and  $|z_0 \sin\theta| < 0.5$~mm for both.
This analysis is available in the {\tt MadAnalysis5}~\cite{Araz:2021akd} Public Analysis Database~\cite{implementationeW}, which has been used along with
{\tt Delphes}~\cite{Delphes} (for detector simulation and object reconstruction)
to compute the expected number of signal events.
The implementation, however, assumes that all leptons originate from the primary vertex and cannot be directly applied for events where $F$ does not decay promptly.
Using the same setup for the MC event generation described above, we obtained samples for promptly decaying $F$'s and used {\tt MadAnalysis5} to apply the analysis selection.
In order to extend the implementation for $c\tau \sim \mathcal{O}({\rm mm})$, 
the weights of the events passing the selection were then re-weighted for a given lifetime value, where the re-weighting factor corresponds to the fraction of $F$ decays taking place within 1~mm of the primary vertex (PV) in the transverse direction and 0.5~mm in the longitudinal direction. 
More concretely, for a given value of proper lifetime ($\tau$), the weight ($w_i$) of a given event is rescaled by:
\begin{equation}\label{eq:lifetimerw}
    w_i(\tau) = w_i \left(1 - e^{-\frac{l_{\rm max}}{ \gamma_1\beta_1 \tau}}\right)\left(1 - e^{-\frac{l_{\rm max}}{\gamma_2 \beta_2 \tau}}\right),
\end{equation}
where $\gamma_i$ is the boost factor for the $i$-th $F$ and $\beta_i$ its corresponding velocity.
In the above expression, $l_{\rm max}$ corresponds to maximum decay length allowed which satisfies the cuts on $d_0$ and $z_0$.
This allows us to approximately reproduce the analysis requirements on the lepton objects.
After computing the signal yields for each signal region and applying the lifetime re-weighting,
limits were once again derived taking into account all 36 signal regions through the 
{\tt Pyhf} statistical model provided by ATLAS and the {\tt Spey} software package.

\section{Results}\label{sec:results}

Assuming the mediator couples universally to all leptons and taking $\lambda_{sh} = 0$, the model described by the Lagrangian of Eq.\eqref{eq:lag} is described by only three free parameters, namely:
\begin{equation}
    m_s, m_F, y_s.
\end{equation}
In addition to these parameters, the dark matter relic abundance can depend on $\TRH$, which is included as a fourth parameter in our numerical analysis.

\subsection{The dark matter relic abundance}\label{sec:DMabundance}

While the dark matter abundance ($\Omega h^2$) depends on all four parameters, the collider phenomenology is independent of the reheating temperature and fairly insensitive to the dark matter mass, unless $F$ and $s$ are mass-degenerate. In this spirit, in what follows we will  present our results in (at least) two different planes, which emphasize different aspects of the model's behaviour: the $(m_s, y_s)$ plane, which focuses on dark matter, as well as the $(m_F, y_s)$ plane which is most relevant for collider searches.
\begin{figure}
    \centering
    \includegraphics[width=0.49\linewidth]{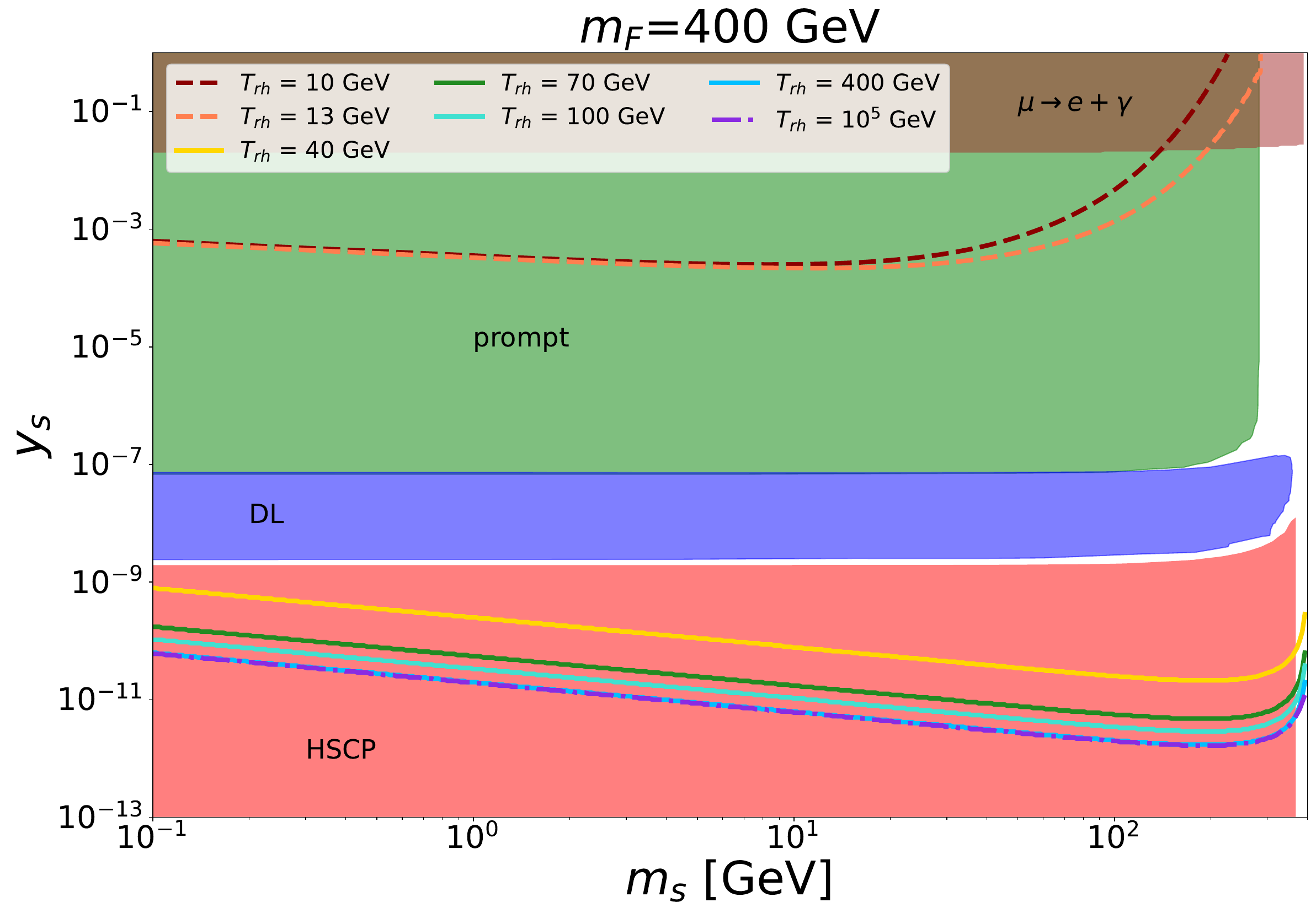}
    \includegraphics[width=0.49\linewidth]{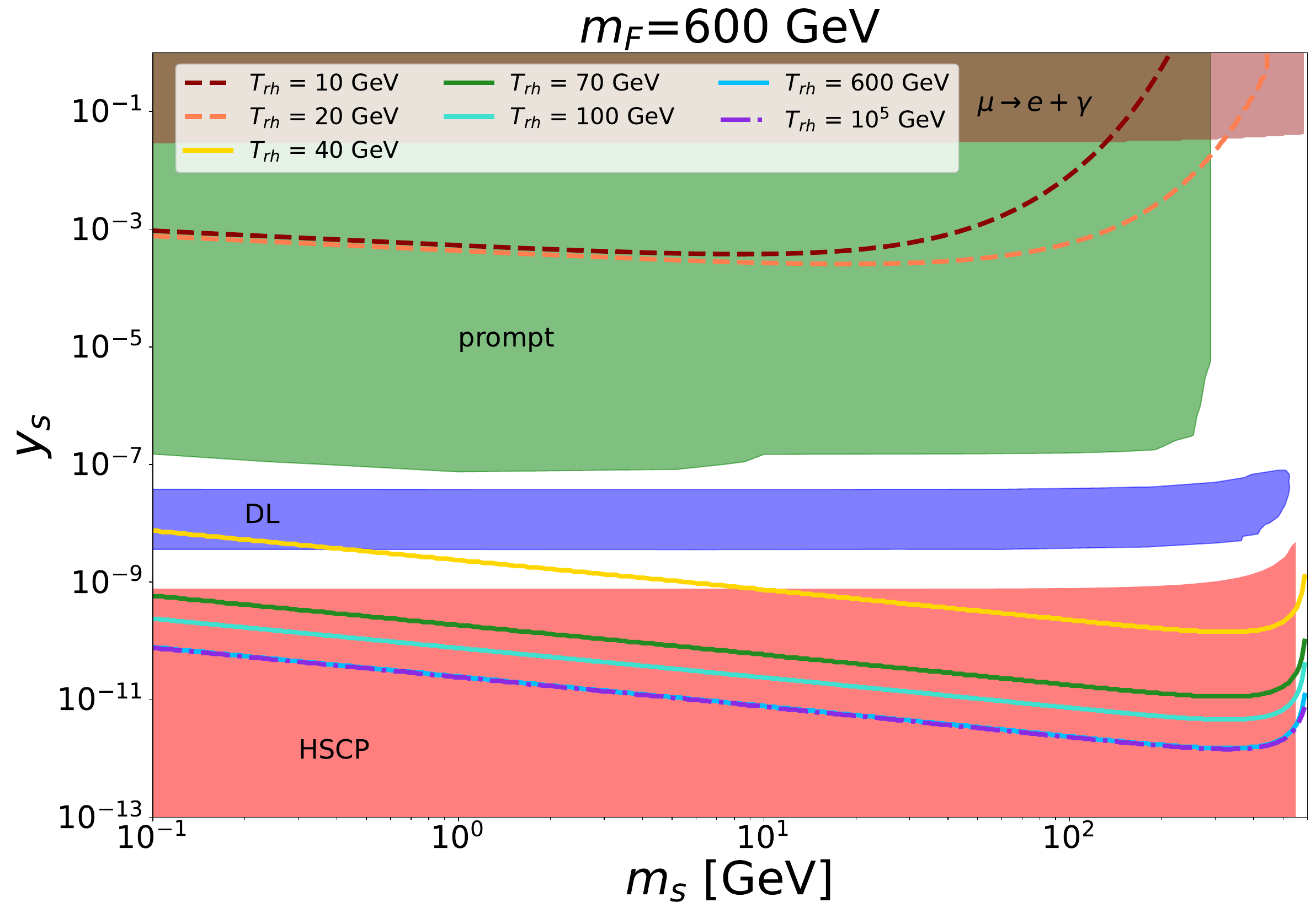} \\ \vspace{0.5cm}
    \includegraphics[width=0.49\linewidth]{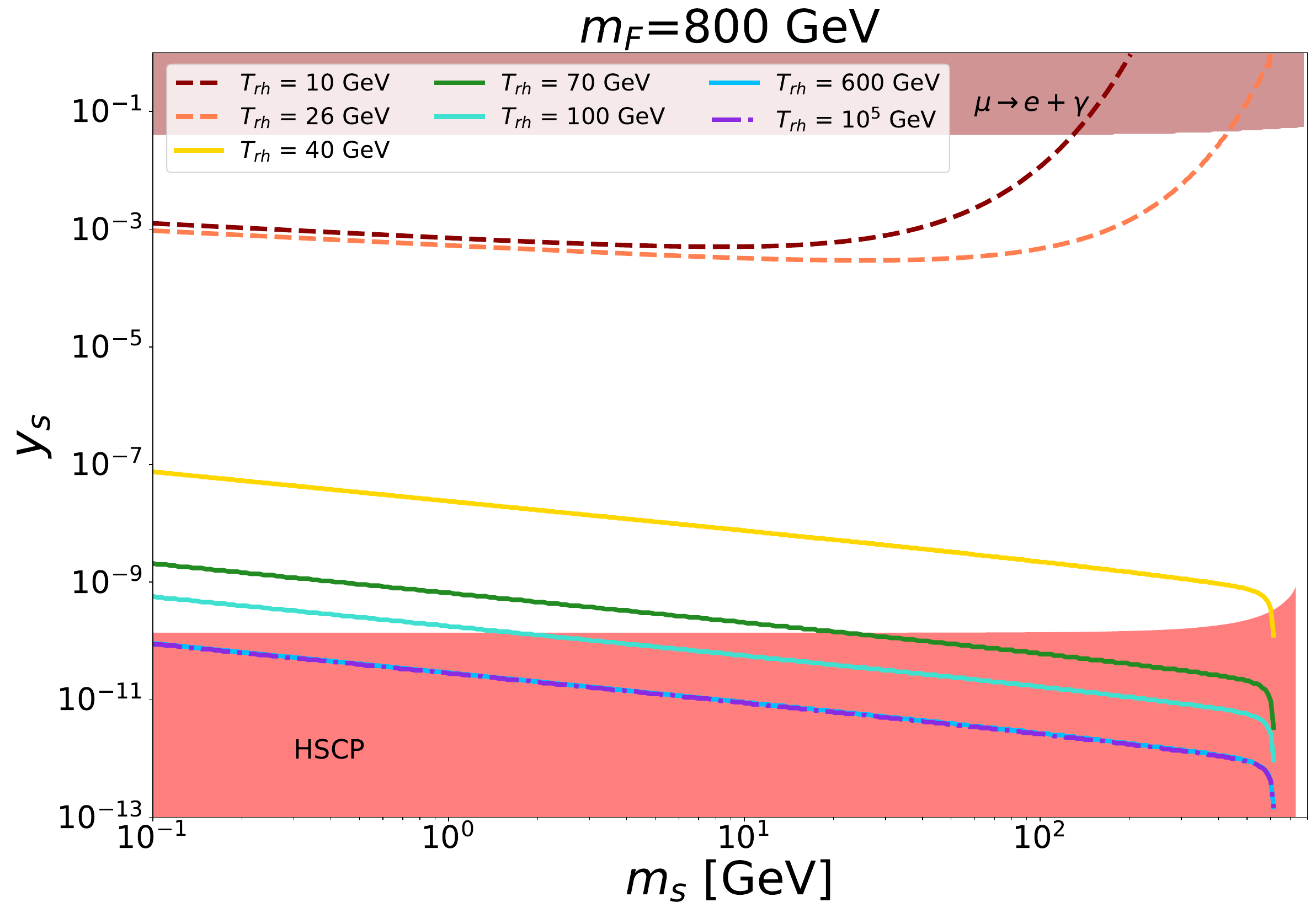}
    \includegraphics[width=0.49\linewidth]{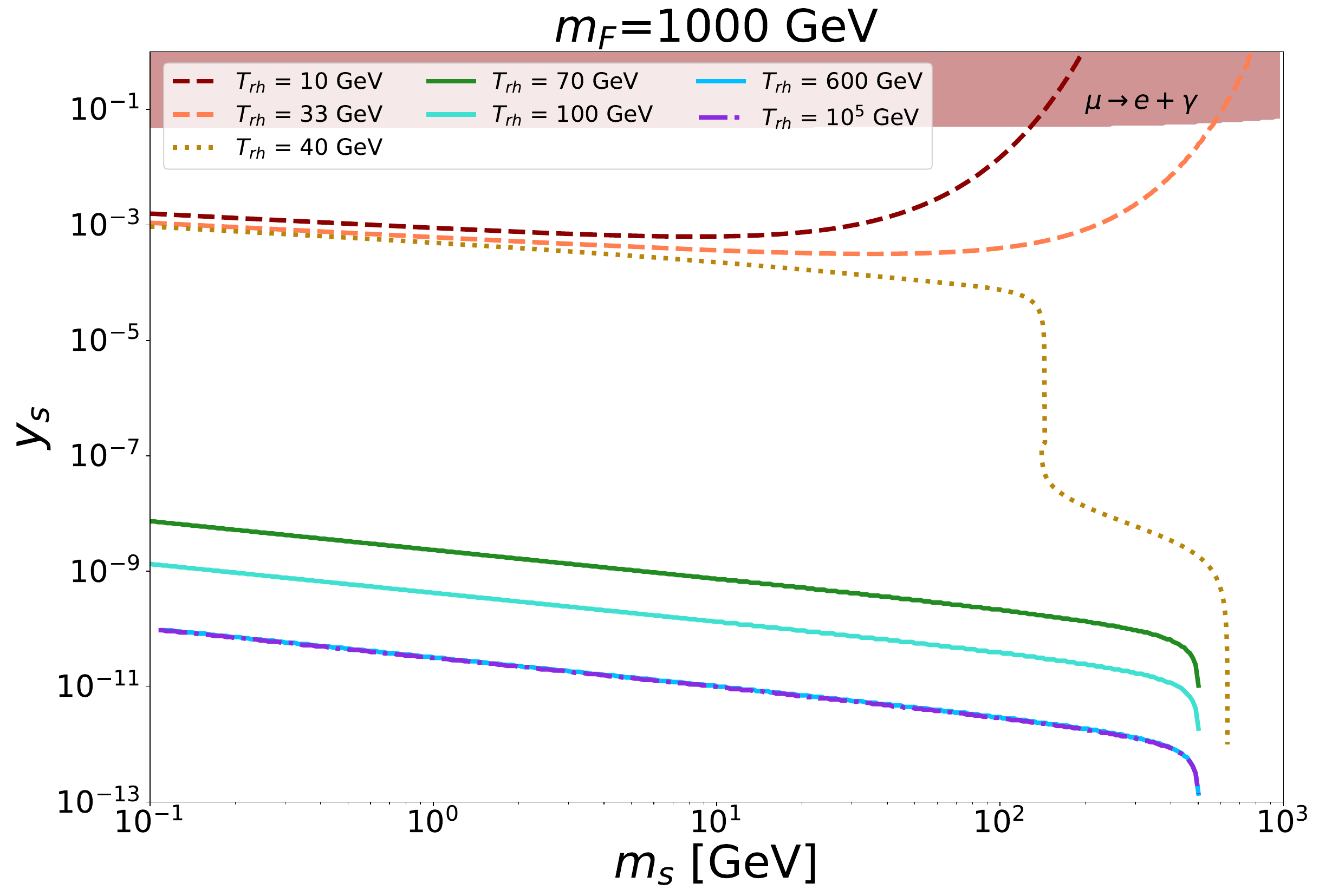}    
    \caption{Contours in the $(m_s, y_s)$ plane along which the relic density constraint is satisfied for different values of the reheating temperature (solid and dashed lines). Each panel corresponds to a different choice of the mediator mass: $m_F = 400$~GeV (top-left), 600 GeV (top-right), 800 GeV (bottom-left) and 1000 GeV (bottom right). The shaded regions correspond to LHC constraints from searches for HSCPs (red), Displaced Leptons (blue) and leptons+MET (green). Constraints from $\mu \rightarrow e\gamma$ are depicted in brown.}
    \label{fig:cosmoplane}
\end{figure}

Starting from the former, in Figure \ref{fig:cosmoplane} we choose four benchmarks for the heavy fermion mass, namely $400$, $600$, $800$ and $1000$ GeV for the top-left, top-right, bottom-left and bottom-right panels, respectively. The contours show the $(m_s, y_s)$ combinations for which the relic density constraint is satisfied for different choices of the reheating temperature. Concerning the latter, we choose: {\it i)} one value corresponding to the IRT regime ($\TRH = 10^5$~GeV, dot-dashed lines); {\it ii)} four values for which the mediator density starts to become Boltzmann suppressed ($\TRH = m_F$ and three values between $m_F$ and the freeze-out temperature $\TFO$, solid lines) and
finally {\it iii)} two values of $\TRH$ lying well bellow $\TFO$ (dashed lines) for all values of $m_F$. The dotted line corresponding to $\TRH = 40$ GeV in the bottom-right panel merits special discussion and we will come back to it shortly afterwards. The LHC constraints from searches for Heavy Stable Charged Particles, Displaced Leptons and leptons + MET are depicted by the red, blue and green shaded areas, respectively. The $\mu \rightarrow e \gamma$ constraint is represented by the brown shaded area towards the upper $y_s$ range of each panel.

In order to understand the behaviour of the cosmologically viable parameter space, let us first point out the fact that, with one exception, in all four panels the different contours are separated into two distinct families: a bottom-most (solid and dot-dashed lines), at $y_s \lesssim 10^{-7}$ and an uppermost (dashed lines), at substantially larger coupling values $y_s \gtrsim 10^{-3}$. This division is due to different processes driving the predicted dark matter relic abundance. 
For the solid curves, the reheating temperature is sufficiently high in order for $F$ to attain thermal equilibrium and freeze-out.
In this case dark matter is mostly produced through $F$ decays as in the standard freeze-in or superWIMP scenarios.
An important distinction, however, is that, for $\TRH < m_F$, the equilibrium density of the mediator becomes Boltzmann suppressed, thus suppressing DM production. As a result the coupling required to achieve the observed relic abundance increases as  $\TRH$ decreases.
Once $\TRH < \TFO$ (dashed lines), the mediator density becomes negligible and the curves fall into the uppermost region of the plane. In this regime dark matter is mostly produced through the annihilation of SM leptons\footnote{Note that at such temperatures, annihilation channels involving vector bosons in the initial state turn out to be sub-leading.} in the thermal bath.
Given the fact that the annihilation is proportional to $(y_s)^4$, the required coupling increases dramatically, reaching rather WIMP-like values, as already shown  in Figure \ref{fig:TRyContours}.

Another interesting aspect of the contours is their dependence on the dark matter mass, $m_s$. 
First we point out that since the relic abundance is proportional to $Y_s m_s$, as $m_s$ increases the yield must decrease.
For small $m_s$ values ($m_s < m_F$ and $m_s < \TRH$), the dark matter production is either controlled by the mediator decay width (in the high $\TRH$/decay regime) or the lepton annihilation cross-section (in the low $\TRH$/annihilation regime). In both cases $Y_s$ is proportional to powers of $y_s$, which must decrease as $m_s$ increases.
As shown in Figure~\ref{fig:cosmoplane}, the decrease is slower in the annihilation region, since the cross-section scales as $y_s^4$, while the decay width scales as $y_s^2$.

For higher $m_s$ values the curves display distinct behaviours, depending on $\TRH$ and $m_F$.
In the decay-dominated regime, where $\TRH > \TFO$, the mediator decay width becomes kinematically suppressed once $m_s \lesssim m_F$, thus decreasing the dark matter production. In this region, the required coupling has to increase to make up for the kinematical suppression, as seen in the bottom-most curves of the upper panels in Figure~\ref{fig:cosmoplane}.
For the lower panels, however, we see the opposite behaviour: as $m_s \gtrsim 500$~GeV, $y_s$ quickly decreases. This is due to the superWIMP contribution which,
as we already mentioned in Section~\ref{sec:themodel}, provides an irreducible contribution to the dark matter relic density and starts to become relevant for high enough $m_s$.
As a result, as $m_s$ increases and the superWIMP contribution becomes dominant, the freeze-in contribution needs to decrease to avoid overproduction of dark matter, which explains the sharp decrease in $y_s$ for high $m_s$.
This only happens for the lower panels, where $m_s$ can reach sufficiently high values. 
Note that, for large enough $m_s$ (and $\TRH$), the superWIMP contribution can already overclose the Universe even for extremely small $y_s$.
Lastly, we comment on the behaviour of the low $\TRH$ curves in the annihilation regime (uppermost curves) as $m_s$ increases.
In this regime, once $m_s \gtrsim \TRH$, the dark matter production from thermal bath scatterings becomes Boltzmann suppressed.
As a consequence, the coupling needs to increase dramatically in order to make up for the kinematical suppression, as seen in the right-most part of the dashed curves in Figure~\ref{fig:cosmoplane}.

\begin{figure}
    \centering
    \includegraphics[width=0.4\linewidth]{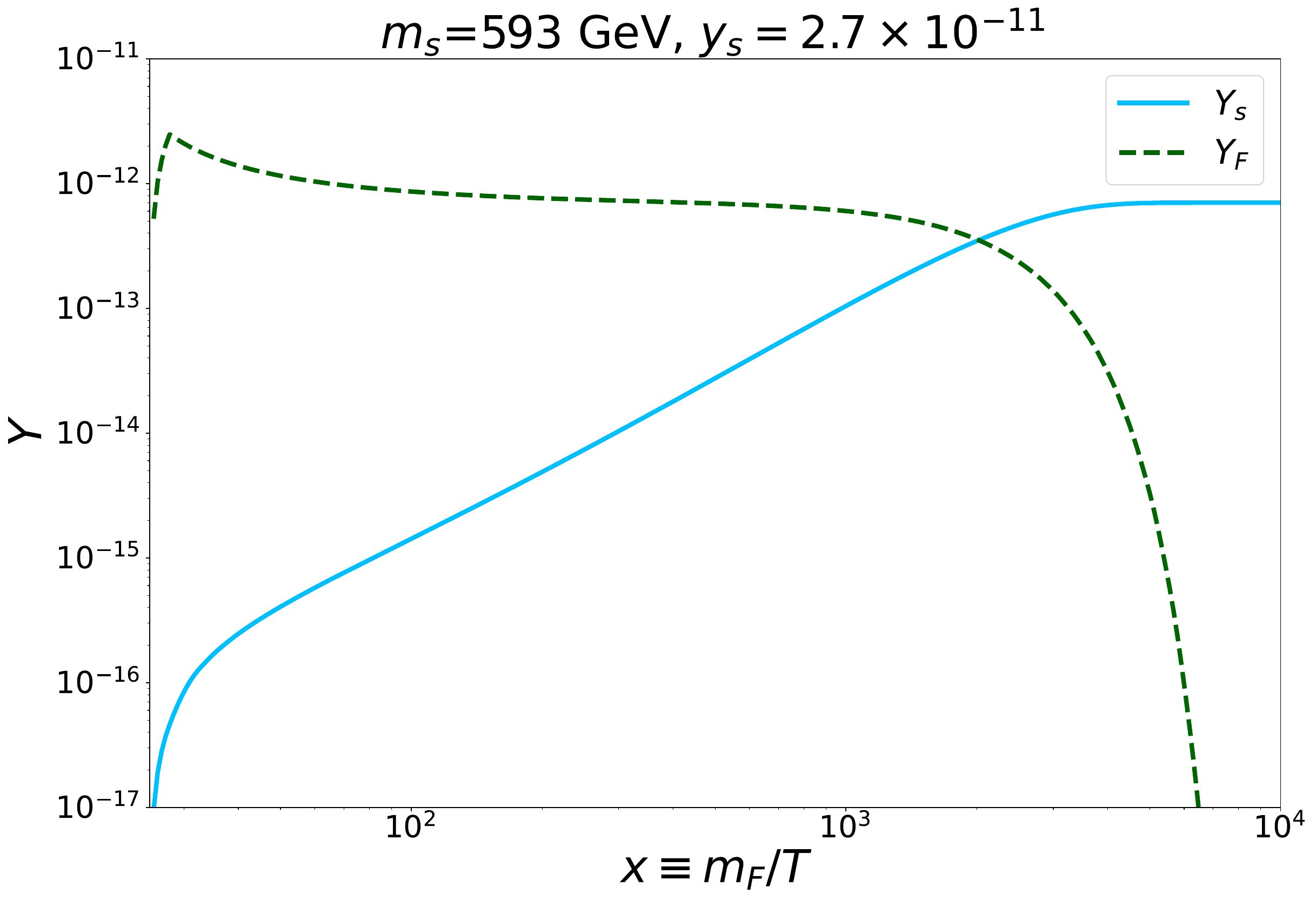}    
    \includegraphics[width=0.4\linewidth]{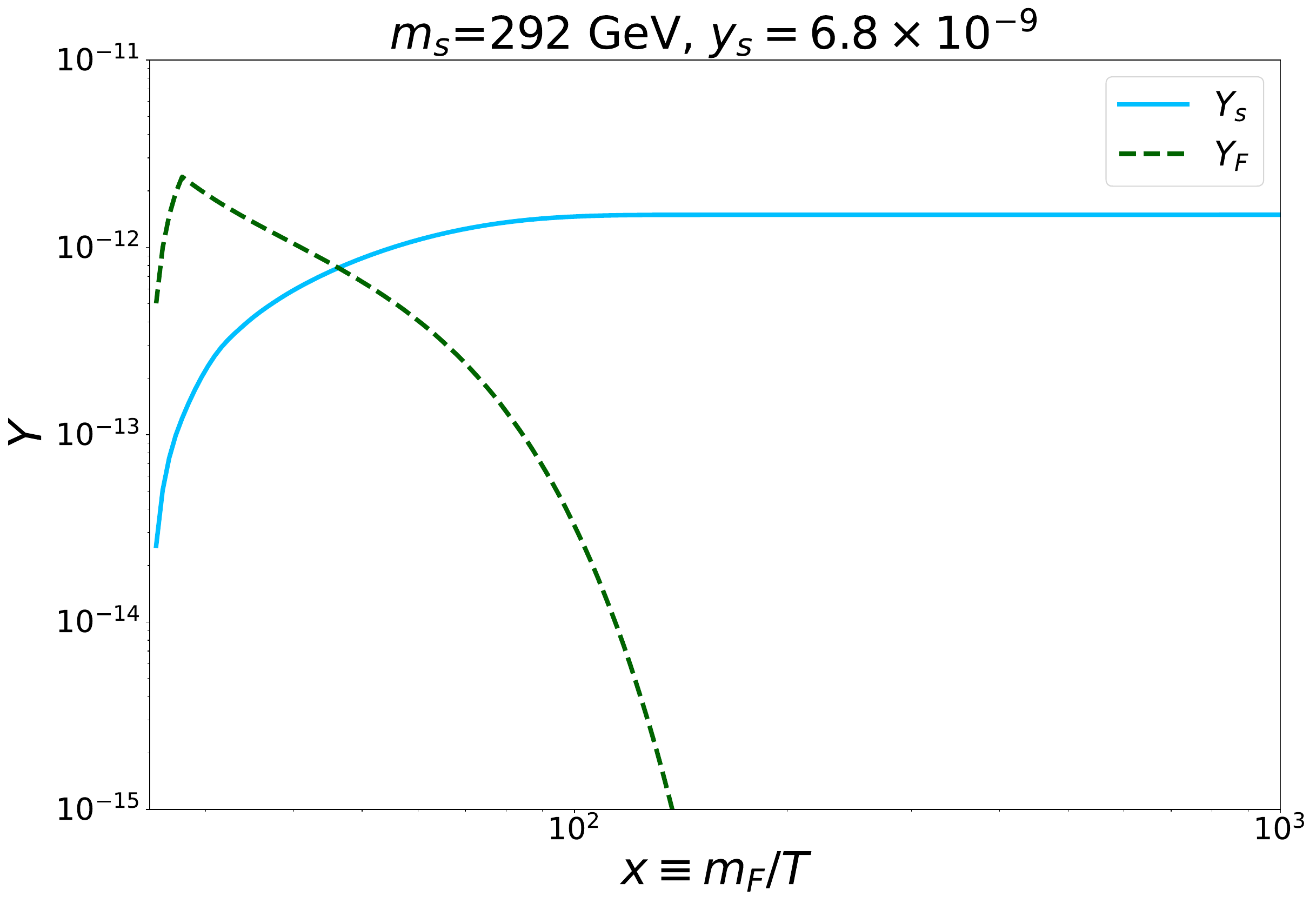} \\
    \includegraphics[width=0.4\linewidth]{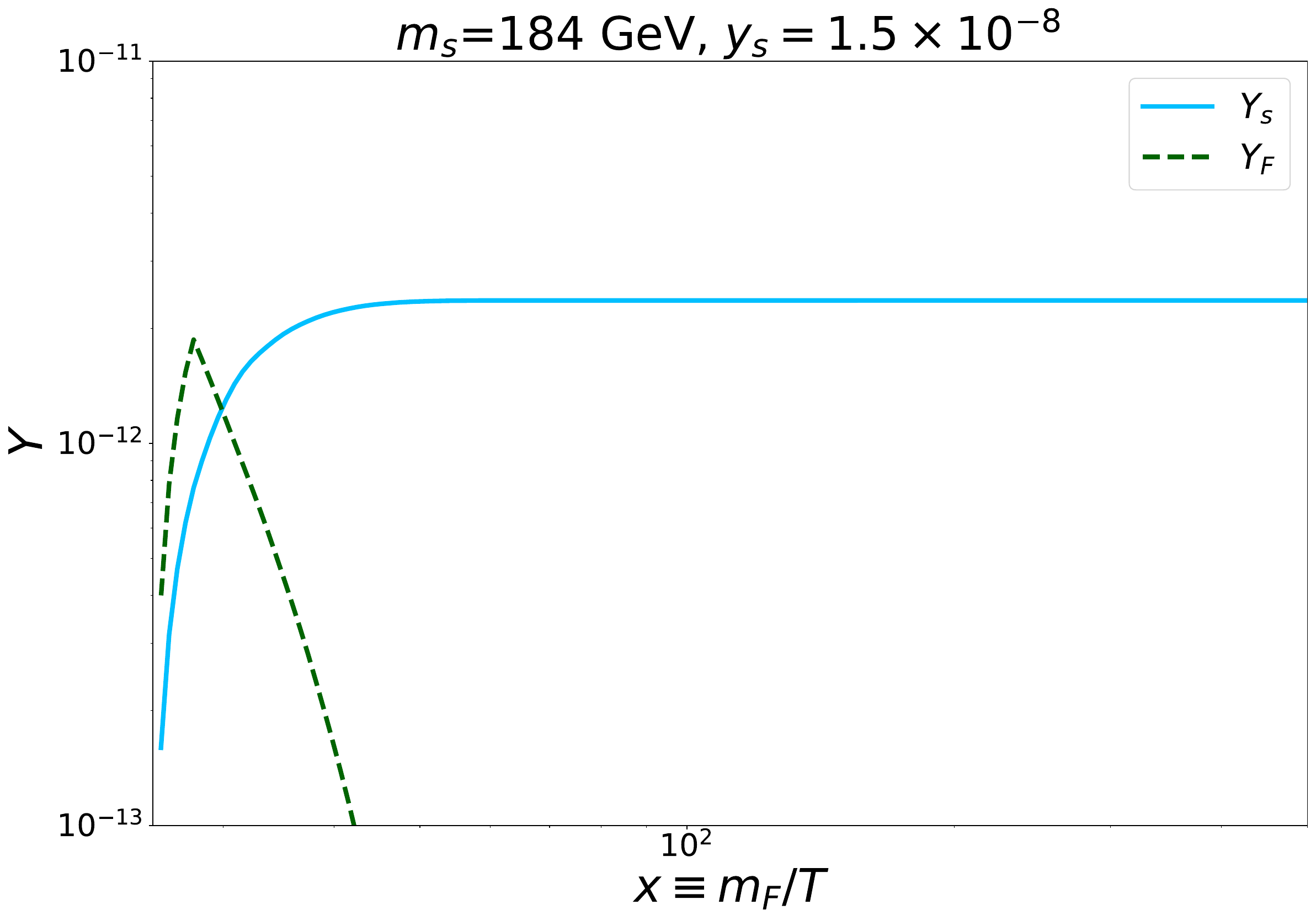}
    \includegraphics[width=0.4\linewidth]{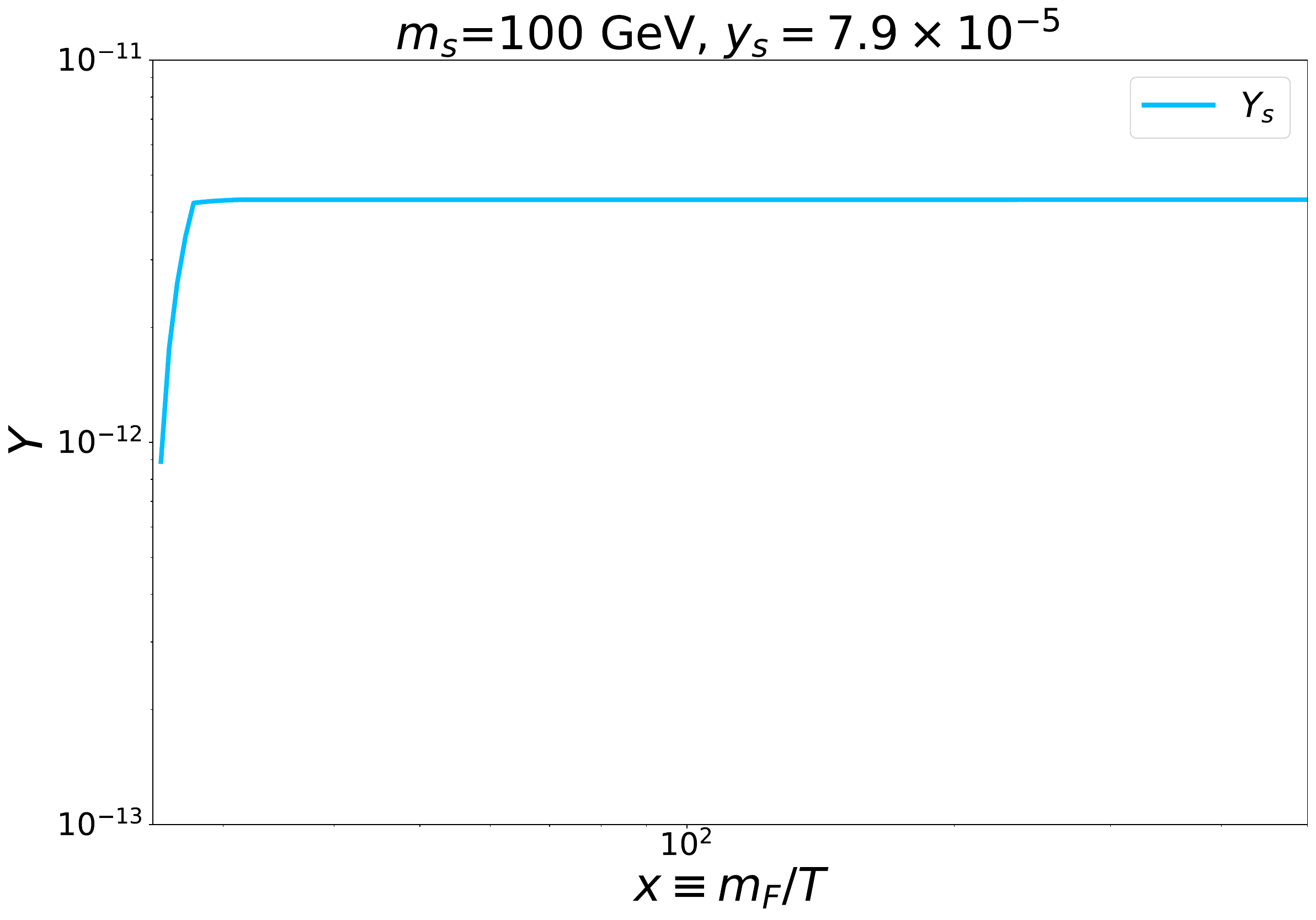}
    \caption{Evolution of the abundances of $s$ (cyan, solid) and $F$ (green, dashed) as a function of $x\equiv m_F/T$ assuming $m_F = 1000$ GeV and $\TRH = 40$ GeV, for four different dark matter masses $m_s = 593$ GeV (top-left), $292$ GeV (top-right), $184$ GeV (bottom-left) and $100$ GeV (bottom-right).}
    \label{fig:combinedevolution}
\end{figure}

Let us now discuss the dotted curve in the bottom-right panel of Figure \ref{fig:cosmoplane}, which corresponds to the benchmark $m_F = 1$ TeV and the scenario $\TRH = 40$~GeV. As we observe, this curve exhibits a transitional behaviour between the lower (solid) and upper (dashed) families of curves; in this specific configuration, the viable parameter space switches from the decay-dominated regime (at high $m_s$) to a scattering-dominated one (for lower $m_s$) around $m_s \sim 170$ GeV. At the largest DM mass values, $m_s \lesssim m_F$, the superWIMP contribution is dominant, requiring extremely small $y_s$ values in order to suppress the freeze-in production.
As $m_s$ decreases, so does the superWIMP contribution and larger couplings are required for the freeze-in (decay-induced) contribution to add up to the observed relic abundance.
As $y_s$ increases, though, a situation similar to the one discussed around Figure \ref{fig:fequilibration} arises: the decay rate of $F$ becomes too large and the mediator never reaches thermal equilibrium. As a result, the mediator production is highly suppressed, requiring even larger $y_s$ values. Eventually, for even smaller $s$ masses (and larger $y_s$ couplings),
the contribution from $F$ decays becomes negligible and SM scattering processes take over.

In order to better illustrate this discussion,  in Figure \ref{fig:combinedevolution} we show the combined evolution of $Y_s$ (blue solid lines) and $Y_F$ (green dashed lines) as a function of $x \equiv m_F/T$ for benchmarks along the $(m_F, \TRH) = (1000, 40)$ GeV line for four values of the dark matter mass: $m_s = 593$ GeV (top-left), $292$ GeV (top-right), $184$ GeV (bottom-left)  and $100$ GeV (bottom-right). As discussed previously, for the first case $F$ follows a standard freeze-out evolution and subsequently decays into $s$, corresponding to the superWIMP-dominated regime. In the second scenario, $F$ still attains equilibrium with the plasma but the decay is much faster. This corresponds to the conventional decay-dominated freeze-in regime. In the last two cases, the abundance of $F$ becomes so suppressed that its evolution becomes irrelevant for dark matter production. 

\begin{figure}
    \centering
    \includegraphics[width=0.8\linewidth]{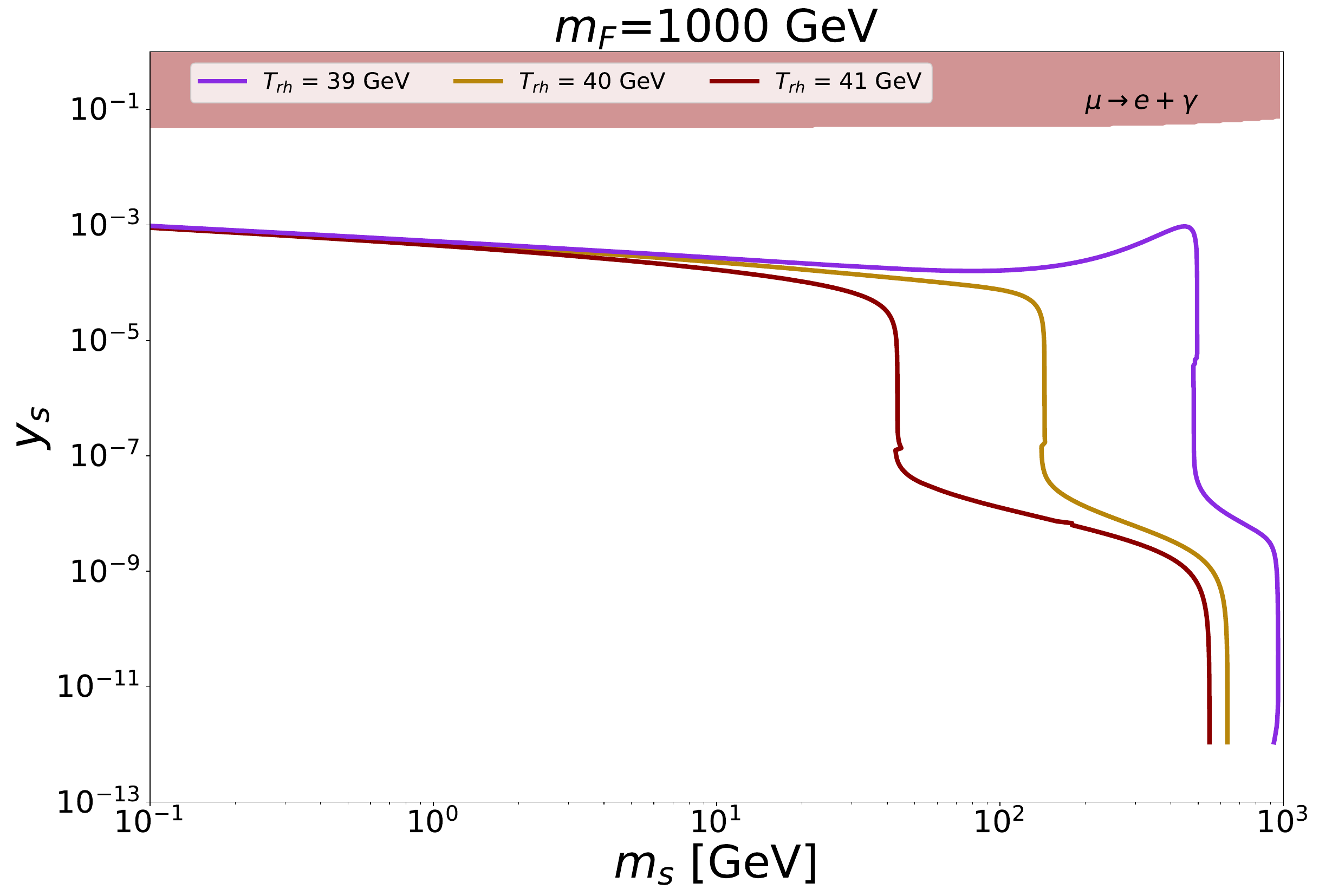}    
    \caption{Contours in the $(m_s, y_s)$ plane along which the relic density constraint is satisfied for $m_F = 1000$ GeV and a few representative values of the reheating temperature close to $T_F^{\rm fo}$.}
    \label{fig:transition}
\end{figure}

The existence of this feature is particularly important because it allows us to understand that \textit{i}) the parameter space between the decay-dominated and the scattering-dominated families of curves is continuously filled and \textit{ii}) how the transition between the two regimes actually occurs. This is illustrated in Figure \ref{fig:transition}, in which we draw contours of $\Omega h^2 = 0.12$ in the $(m_s, y_s)$ plane assuming $m_F = 1000$ GeV and for a few representative $\TRH$ values close to $T_F^{\rm fo} \simeq 38.5$ GeV. As we observe, as $\TRH$ decreases, the transition regime moves to lower and lower DM masses. And, in the limit of a continuous variation of the reheating temperature, the entire parameter space is filled up. In passing, let us also note that as $\TRH$ decreases, the superWIMP-dominated regime moves to higher DM masses due to the gradual suppression of the ``would-be-asymptotic'' abundance of $F$, \textit{cf} Figure \ref{fig:fevolution}.

\subsection{Collider and Flavour Constraints}\label{sec:phenoresults}

Moving to LHC constraints, by inspecting Figures \ref{fig:lifetimes} and \ref{fig:cosmoplane} we see that for sufficiently large $\TRH$, where dark matter production is dominated by $F$ decays, we have $y_s < 10^{-9}$, leading to decay lengths of the order of a few meters or higher for the mediator.
As a result, a large fraction of the mediators produced at the LHC mostly decays outside the ATLAS and CMS detectors. In this regime the HSCP searches are the most constraining ones and independent of $m_s$ as long as $m_s \ll m_F$. Once $m_s \sim m_F$, the decay becomes kinematically suppressed and large lifetimes can still be achieved even for $y_s \sim 10^{-8}$. As a result, the HSCP exclusion reaches higher $y_s$ values in this compressed region.
This is shown by the red bands in Figure~\ref{fig:cosmoplane}, which can exclude mediator masses up to $m_F \lesssim 800$~GeV.\footnote{For the highly compressed region ($\Delta m \lesssim$~1 GeV), constraints from disappearing track searches can become relevant. However, using the results from Refs.\cite{ATLAS:2022rme,CMS:2020atg}, we have estimated that only mediator masses up to $m_F \sim 400$~GeV would be excluded. This region is almost entirely excluded by the other searches considered here.}
For larger $m_F$ values the mediator production cross-section falls below 0.1~fb, resulting in too few signal events for the searches to exclude the model.
Once $y_s \gtrsim  10^{-8}$, the mediator decay length becomes of the order of a few centimeters, as shown in Figure~\ref{fig:lifetimes}. 
For such lifetimes, $F$ decays primarily within the tracker and the search for displaced leptons (DL) becomes the most sensitive one, as illustrated by the blue regions shown in Figure~\ref{fig:cosmoplane}.
This search, however, requires leptons to be produced within the tracker and quickly loses its sensitivity once the decay length is below a few millimeters.
As a result, the DL search is able to exclude the region $2\times 10^{-9} < y_s < 7 \times 10^{-8}$ ($10^{-8} < y_s < 3 \times 10^{-8}$) for $m_F = 400$~GeV ($m_F = 600$~GeV), while for $m_F = 800$~GeV and 1~TeV the mediator production cross-section falls below the search sensitivity.\footnote{There is a very small region excluded by the DL search for $m_F = 800$~GeV and $m_s \simeq 400$~GeV. However, since it is barely visible, we have chosen not to display it in Figure~\ref{fig:cosmoplane}.}
We see that this region of parameter space can explain the observed relic abundance for intermediate $\TRH$ values, where the mediator production becomes Boltzmann-suppressed.
As discussed above, for even smaller reheating temperatures, dark matter production takes place in the annihilation-dominated regime, which requires $y_s \gtrsim 10^{-3}$. In this case the mediator lifetimes are well below a millimeter, leading to prompt decays.
Searches for di-leptons and missing energy are then the most sensitive ones and are able to exclude mediator masses up to $m_F = 650$~GeV.
The exclusion bounds are depicted by the green regions in the top panels in Figure~\ref{fig:cosmoplane}, while for the panels with $m_F = 800$~GeV and 1~TeV the cross-sections are too small to be excluded by prompt searches.

Moving on to non-collider bounds, searches for $\mu \rightarrow e\gamma$ decays constrain coupling values in the ${\cal{O}}(10^{-2})$ ballpark. When the mediator mass is relatively small, these constraints are sub-leading with respect to those stemming from LHC searches. However, and as expected, they are capable of probing larger values of $m_F$ and $m_s$. In this sense, they provide highly complementary constraints on our model.

In summary, as we can see in Figure~\ref{fig:cosmoplane}, a large region of the model parameter space is consistent with the observed relic abundance for a given mediator mass, if appropriate values of $\TRH$ and $m_s$ are chosen. However, these quantities have little impact on the LHC constraints. For this reason, it is interesting to also display the collider constraints in the $(m_F,y_s)$ plane, since these are the two parameters that are the most relevant ones for collider phenomenology. 
\begin{figure}
    \centering
    \includegraphics[width=0.9\linewidth]{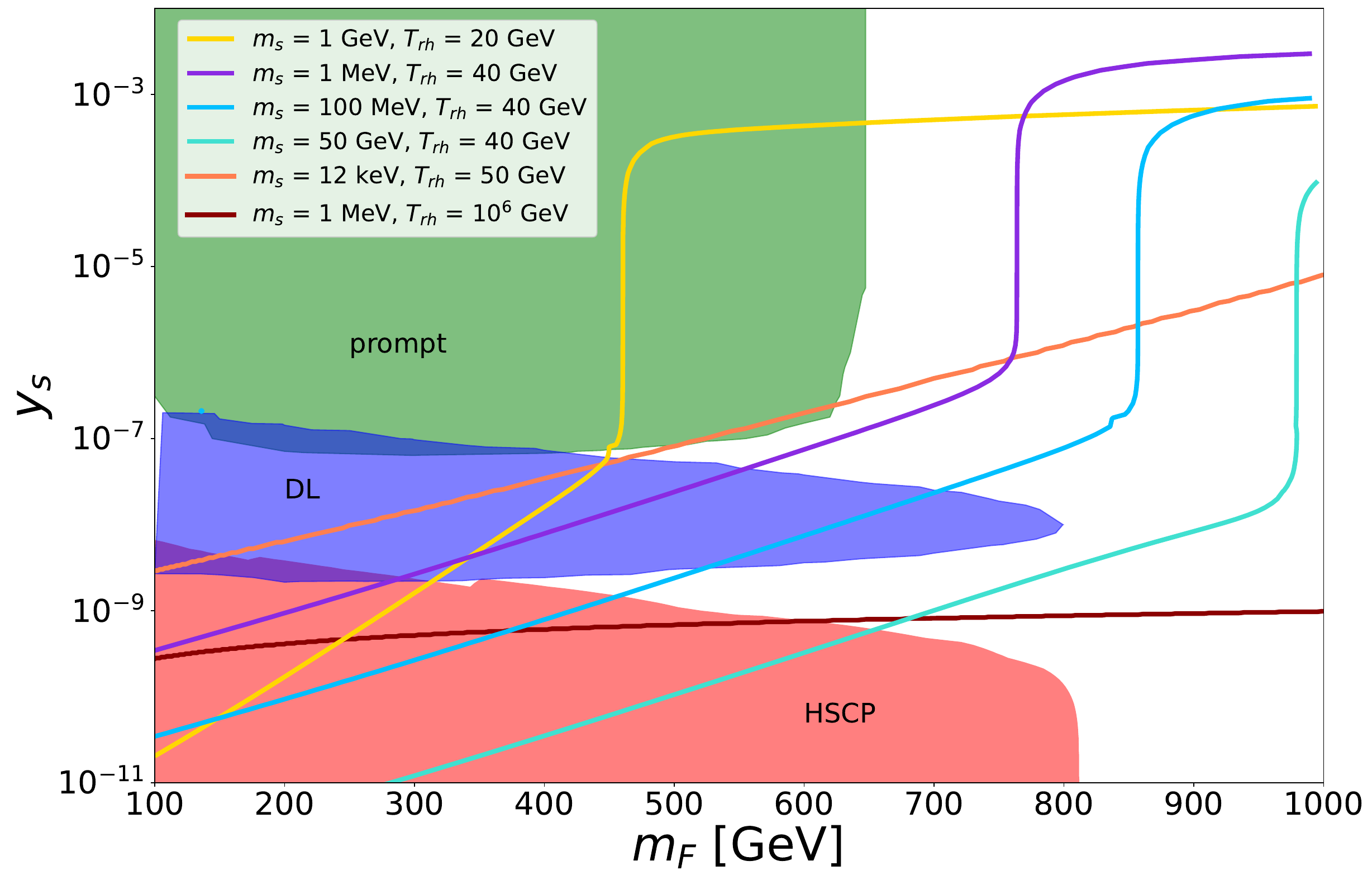}
    \caption{LHC constraints from searches for Heavy Stable Charged Particles, Displaced Leptons and prompt leptons accompanied by missing transverse energy (red, blue and green shaded regions, respectively). The relic density constraint can be satisfied for different choices of $m_s$ and $\TRH$ as shown in the legend.}
    \label{fig:colliderplane}
\end{figure}

In Figure~\ref{fig:colliderplane} we show the regions excluded by the HSCP, displaced lepton and prompt searches for $m_s \ll m_F$ (red, blue and green shaded regions, respectively). In addition, we display contour curves satisfying the observed relic density for several choices of $m_s$ and $T_{\rm rh}$. The brown curve ($m_s = 1$ MeV, $T_{\rm rh} = 10^6$ GeV) corresponds to a standard, relatively low DM mass and effectively infinite reheating temperature scenario, in which DM is produced in the IR through $F$ decays. The same mechanism drives the DM abundance along the orange curve ($m_s = 12$ keV, $T_{\rm rh} = 50$ GeV), albeit at larger coupling due to \textit{i}) the Boltzmann suppression of the $F$ abundance and \textit{ii}) the lower DM mass. The remaining curves exhibit the ``regime-flip'' behaviour described previously, as the leading contribution to the dark matter abundance transitions from a decay-dominated regime (lower $m_F$ values) to a scattering-dominated one (higher $m_F$ values), with the transition point depending on $T_{\rm rh}$.

As we can see, the HSCP and DL searches have the potential to exclude mediator masses up to 800~GeV, while the prompt search excludes values up to $m_F = 650$~GeV. LLP searches can test mostly decay-dominated freeze-in scenarios. Prompt searches, on the other hand, become relevant for scattering-dominated ones, very low DM masses (combined with moderate reheating temperatures) or, eventually, the transition regime. It is also interesting to notice that for $y_s \simeq 10^{-7}$ and $y_s \simeq 10^{-8}$ the searches have some overlap, which shows their complementarity and good coverage of the parameter space.

All in all, Figures \ref{fig:cosmoplane} and \ref{fig:colliderplane} show how different types of collider searches can probe different regions of parameter space which become viable under different cosmological assumptions. As a rule of thumb, conventional (high reheating temperature) freeze-in scenarios fall mostly within the reach of searches for Heavy Stable Charged Particles. The only exception concerns the case in which dark matter is very light, close to its lowest allowed value by Lyman-alpha forest observations \cite{Belanger:2018sti}. In this case, the required coupling values lead to mediator lifetimes which are more relevant for displaced lepton searches. As the reheating temperature becomes smaller, and in particular when it becomes smaller than the mediator mass, the necessary coupling increases and the mediator becomes shorter-lived.

As a final comment let us stress that, as already mentioned, throughout our analysis we assumed a universal coupling $y_s$ between dark matter, the vector-like lepton and all lepton flavours.
If, instead, we had assumed a Minimal Flavour Violation (MFV) scenario, with a coupling structure $\y \propto m_{l^\alpha}$, $F$ would mostly couple to $\tau$'s. In this case we do not expect the behaviour of the dark matter relic density to significantly differ from the universal scenario. The main difference is that since only one leptonic channel (either for decays or scatterings) is relevant, slightly larger $y_s$ values would be required. From the point of view of collider constraints, however, the results could change considerably. In particular, searches for final states with $\tau$'s typically have larger backgrounds or smaller signal efficiencies (due to $\tau$-tagging issues), which result in weaker collider constraints. For instance, the displaced lepton search from Ref.~\cite{ATLAS:2020wjh} excludes long-lived smuons up to 700~GeV, while long-lived staus are only excluded up to 325~GeV.

Furthermore, in a MFV-like scenario, the couplings of $F$ (and $s$, for that matter) to light leptons would be suppressed and the constraints from $\mu \rightarrow e\gamma$ would become less relevant.
Nonetheless, the flavor constraints presented in Figure \ref{fig:cosmoplane} can be promptly translated into \textit{any} scenario with the same particle content, as the $\mu \rightarrow e\gamma$ prediction depends simply on the product $y_s^e \times y_s^\mu$.

\section{Conclusions}\label{sec:conclusions}

In this paper we performed a comprehensive analysis of a ``charged parent'' freeze-in dark matter model \cite{Belanger:2018sti} with particular focus on scenarios where,
following inflation, the Universe (instantaneously) reheated to a temperature which is comparable-or-lower than the relevant mass scales. This leads to Boltzmann-suppressed dark matter production \cite{Belanger:2018sti,Brooijmans:2020yij} and opens up the possibility of freeze-in occurring for strong microscopic couplings between DM and the SM thermal bath \cite{Brooijmans:2020yij, Cosme:2023xpa}.

As we showed, and depending on the temperature of the Universe at the onset of radiation domination, in such configurations the predicted dark matter relic abundance can depend highly non-trivially on all parameters of the microscopic model. Moreover, tracking the number density evolution of several particle species may be critical for a realistic evaluation of the predicted dark matter relic density: unexpected behaviours may appear across the model's parameter space, some of which we (to the best of our knowledge, for the first time) pointed out. In particular, we showed that there is a continuous transition between decay-dominated and scattering-dominated freeze-in, which is reminiscent of the smooth transition between freeze-in and freeze-out \cite{Cosme:2023xpa,Arcadi:2024wwg}.

Our analysis highlights the fact that there is no clear-cut distinction between microscopic models of dark matter, thermal histories and, hence, dark matter production mechanisms. Assessing the cosmologically viable parameter space of concrete dark matter models can depend heavily on the underlying assumptions concerning the thermal history of the Universe. Moreover, as we saw in many occasions it may become necessary to simultaneously track the number density evolution of multiple particle species in order to capture the underlying physics and extract reliable predictions.

Besides, these comments are not just relevant for cosmology; they can have important consequences for phenomenology as well. In particular, as we showed, modified cosmological assumptions can lead to different experimental searches and/or constraints becoming relevant in different regions of each microscopic model's cosmologically viable parameter space. In this sense, different experimental searches such as the ones we studied in this paper are highly complementary: they do not just probe ``dark matter models''. Instead, they probe combinations of microscopic models and thermal histories.

\section*{Acknowledgments}\label{sec:Acknowledgments}

AG and TR acknowledge fruitful discussions with G. B\'elanger and S. Pukhov.
AL is supported by FAPESP under grants no.~2018/25225-9 and 2021/01089-1 and CNPq grant no.~300278/2025-0. LMDR acknowledges support from CNPq under grant no.~140343/2022-9.
\newpage

\bibliographystyle{JHEP}
\bibliography{references}
\end{document}